%% file: main.tex
\documentclass[final,5p,times,twocolumn,authoryear]{elsarticle}
\pdfoutput=1

\usepackage{hyperref}
\hypersetup{
    colorlinks=true,
    linkcolor=myblue,
    citecolor=myblue,
    urlcolor=myblue
}
\usepackage[pagewise]{lineno}
\usepackage{newtxtext,newtxmath}
\usepackage{xcolor}
\usepackage{multirow}
\usepackage{threeparttable}
\usepackage{subcaption}
\usepackage{mwe}
\usepackage{comment}
\usepackage{adjustbox}
\usepackage{graphicx}
\usepackage{amsmath}
\usepackage{footnote}
\makesavenoteenv{table}
\makesavenoteenv{tabular}
\usepackage{tabularx}
\usepackage{graphicx}
\usepackage{soul}
\usepackage{epstopdf}

\usepackage{float}
\usepackage{lipsum}
\usepackage{pdfpages}

\usepackage[utf8]{inputenc} 
\usepackage[T1]{fontenc} 

\newcommand{\src}{4U 1630--472 }

\begin{document}
\date{Accepted XXX. Received YYY; in original form ZZZ}
{
\makeatletter
\renewcommand*{\corref}[1]{\@gobble}
\providecommand*{\cnotenum}[1]{}
\makeatother

\begin{frontmatter}
\journal{Journal of High Energy Astrophysics}
\title{Spectro-temporal Investigation of Quasi-periodic Oscillations From Black Hole X-ray Binary 4U 1630-472 Using \emph{NICER}}

\author[1,2,3]{Ansh Chopra\corref{cor1}}
\ead{anshchopra2@gmail.com}
\author[1]{Manoneeta Chakraborty}
\author[4]{Unnati Kashyap}

\address[1]{Department of Astronomy, Astrophysics, and Space Engineering, Indian Institute of Technology Indore, Indore 453552, India.}
\address[2]{Department of Astroparticle Physics, Gran Sasso Science Institute, Viale F. Crispi 7,I-67100,L’Aquila (AQ), Italy.}
\address[3]{INFN – Laboratori Nazionali del Gran Sasso, I-67100, L’Aquila (AQ), Italy.}
\address[4]{Physics \& Astronomy Department, Texas Tech University, Lubbock, USA.}

\begin{abstract}
We present a comprehensive analysis of the spectro-temporal characteristics of the X-ray variabilities from black hole X-ray binary \src during its three outbursts (2018, 2020, and 2021) as observed by \emph{NICER}. 
We detected 27 Quasi-Periodic Oscillations (QPOs), out of which 25 were observed during the 2021 outburst. In this study, we specifically focus on the relationship between spectral and timing parameters and the frequency of type-C QPOs in the 2021 outburst of the black hole binary \src during its rising phase. We found strong correlations between the photon index of the non-thermal emission and the QPO frequency. We also observed a critical frequency at $\sim$ 2.31 Hz, above which the behavior of the Q-factor of the QPO changed significantly with the QPO frequency. We further identified two events characterized by a surge in the total flux, corresponding to the disappearance of type-C QPOs. Although the first event appeared like an X-ray flare, during the second event, the source reached a state with a total flux higher than 10$^{-8}$ erg/cm$^{2}$/s and exhibited a different type of QPO with lower frequencies and weaker amplitudes. We compare our results with the previously reported QPO characteristics for black hole outbursts and discuss the various models that could interpret the critical frequency and potentially 
explain the origin and evolution of these type-C QPOs.
\end{abstract}


\end{frontmatter}
}

\input{introduction}
\input{observations}
\input{results}
\input{discussion}
\input{conclusion}

\section*{Acknowledgements}
This work has used software and data provided by the \emph{NICER} team and the High Energy Astrophysics Science Archive Research Center (HEASARC), a service of the Astrophysics Science Division at NASA/GSFC. We acknowledge support for the same.

\bibliographystyle{elsarticle-harv}
\bibliography{references.bib}
\label{lastpage}
\end{document}

%% file: introduction.tex
\section{Introduction} \label{introduction}

Black hole X-ray binaries (BHXBs) are intriguing astrophysical systems comprised of a black hole and a companion star. After spending most of their time in quiescence, black hole low-mass X-ray binaries (BH LMXBs) undergo outbursts when the black hole actively accretes matter from the low-mass companion star through the Roche lobe overflow, leading to a significant increase in luminosity \citep{Remillard2006}. These sporadic outbursts may last from several weeks to months. During the rising phase of an outburst, BHXBs initially appear in the Low/Hard state (LHS) and, as the outburst progresses, move to the High/Soft state (HSS) via the Hard-Intermediate state (HIMS) and Soft-Intermediate state (SIMS),  where the distinct spectral states are characterized by the spectral hardness and intensity \citep{belloni2005}. Finally, before returning to the quiescence state, BHXBs transition back to the LHS. This evolution can be traced by the Hardness Intensity Diagram (HID), which shows the count rate of the source with the Hardness Ratio (HR), which is the ratio of the count rates in the hard and soft X-ray bands. A `q' shaped profile is usually expected in the case of BHXBs \citep{motta2011, belloni2005}. 
However, certain BHXBs were observed to exhibit atypical `c' like tracks in the HID during a few of their outbursts \citep{Baby}.

During the outbursts, transient BHXBs exhibit a wide range of spectral and timing variations \citep{klis2004, klis1988}. The X-ray spectrum of a BHXB usually consists of a thermal and a non-thermal component dominating in the HSS and LHS, respectively. The former is believed to come from a geometrically thin and optically thick accretion disk \citep{shakura}, while the latter is proposed to be produced by the Componization of soft photons by an extended cloud made of plasma called ``corona'' \citep{sunyaev}. 

Understanding the temporal behaviour of BHXBs plays a pivotal role in deciphering the underlying processes and geometrical configurations of their accretion flows \citep{belloni&stella2014}. Power spectral analysis emerges as a valuable tool for probing X-ray variability in BHXBs, revealing intriguing phenomena such as Low-Frequency Quasi-Periodic Oscillations (LFQPOs) \citep{zhang2023, zhang2022, belloni&tomaso}. LFQPOs, characterized by central frequencies spanning 0.1 to 30 Hz, exhibit diverse characteristics depending on their centroid frequency, time/phase-lag, and amplitude \citep{ingram2019}. The fractional RMS (root-mean-square) of the LFQPOs are, in general, higher during the rising part of the outburst \citep{dieters}. The presence and features of LFQPOs exhibit correlations with the spectral states of BHXBs \citep{homan2005}. Classifiable by their attributes, LFQPOs encompass type-C, the predominant and robust QPOs with fractional RMS amplitude reaching up to 15\%, typically encountered in the LHS and HIMS; type-B and type-A QPOs, prevalent in the SIMS and HIMS. Instances of transitions between distinct QPO types during hard-to-soft transitions have been documented \citep{Ma2023, motta2011, zhang2021, sriram2013}. It has been postulated that a precessing hot inner flow/jet base may modulate X-ray emission, giving rise to type-C QPOs (\cite{ingram2009}; \cite{ingram2019} and references therein). Nevertheless, a comprehensive model for type-A and type-B QPOs remains elusive.

Alongside LFQPOs, quasi-regular modulations (QRMs) with long periods of $\sim$10-20 s have been observed in the light curves of the BHXB \src. These QRMs were first observed in \src during its 1998 outburst by \citet{trudolyubov2001} and recently \citet{yang2022} reported QRMs at 60 mHz in the 2021 outburst. \citet{yang2022} also reported that the fractional RMS of the QRMs increased with photon energy. During both the outbursts of this source in 1998 and 2021, these QRMs were seen to appear in a narrow flux range of $\sim$ $1.4 \times 10^{-8}$ erg s$^{-1}$ cm$^{-2}$ in the 3-20 keV range thereby, showing a peculiar dependence on the accretion rate. For BHXBs like GRS 1915+015, similar modulations in the light curve with frequencies in the mHz range have been observed and are believed to be associated with evaporation or ejection near the inner accretion disc, driven by the radiation pressure \citep{neilsen2011}.

The energy dependence of frequency and amplitude of the LFQPOs can offer key insight into the understanding of the accretion processes around black holes.
It has been reported that the fractional RMS amplitude of LFQPO increases with energy up to a certain threshold, beyond which it flattens or exhibits variable behaviour \citep{Qu, Huang}. The properties of the QPO have been found to evolve as the source goes through the duration of an outburst, transitioning between the different states. This behaviour points towards a connection between the radiative processes and the QPO generation. In this regard, it is crucial to examine the role the spectral parameters play in the generation of this distinctive temporal feature.

The BHXB 4U 1630--472 was discovered in 1969 by Vela 5B satellite \citep{Priedhorsky1986}, and since its discovery, this source has exhibited numerous outbursts, displaying a wide range of behavior of its timing and spectral properties \citep{parmar1995, capitanio}. \citet{capitanio} also argued that the periodic outbursts observed in \src may be attributed to the gravitational influence of a third body in orbit around the system. It is a highly absorbed source with the neutral Hydrogen column density as $N_{H} = (4-12) \times 10^{22}$ atoms cm$^{-2}$ \citep{kuulkers1998}. \citet{Cui} reports an N$_{H}$ as high as 16$\times 10^{22}$ cm$^{-2}$ for \src. \citet{Kalemchi} estimated the distance of \src to be between 4.7$\pm$0.3 kpc and 11.5$\pm$0.3 kpc whereas \citet{capitanio}, \citet{augusteijn2001}, and \citet{seifina2014} argued for the distance of the source to be in the range of 10--11 kpc.
While the optical or infrared counterpart of 4U 1630--472 remains unidentified, the mass of the black hole was estimated as $\sim$10 \(M_\odot\) \citep{seifina2014}, while \citet{king2014} reported an inclination angle of $i = 64^{+2}_{-3}$ degrees, and a spin parameter $a_{*} = 0.985^{+0.005}_{-0.014}$ through various techniques. During the recent outburst of \src in 2022, \citet{Ankur} estimated the spin parameter $a_{*} \sim 0.93$ during the HSS observations, further supporting that \src contains a highly spinning black hole.

In this study, we present detailed spectral and timing analyses focusing on the QPO characteristics detected during three outbursts of 4U 1630--472, namely the 2018, 2020, and 2021 outbursts, using high time-resolution data from the Neutron Star Interior Composition ExploreR (\emph{NICER}) telescope. 

By examining the variability and energy dependencies of LFQPOs, aided by the exceptional capabilities of \emph{NICER}, as well as the correlation of LFQPOs with the timing and spectral parameters, we aim to enhance our understanding of the accretion processes and associated phenomena in BHXBs, especially with regard to the origin and evolution of rapid variability processes. 
In Section~\ref{sec:obs}, we describe the observations considered for this study and the corresponding data reduction process. The analysis and results covering the spectral analysis and the variability studies are reported in Section~\ref{sec:results}. Finally, we discuss the interpretations of the results in Section~\ref{discussion}. 

%% file: observations.tex
\renewcommand*{\corref}[1]{\@gobble}
\providecommand*{\cnotenum}[1]{}
\section{OBSERVATIONS AND DATA REDUCTION}\label{sec:obs}

\begin{figure*}
  \centering
  \includegraphics[width=\textwidth]{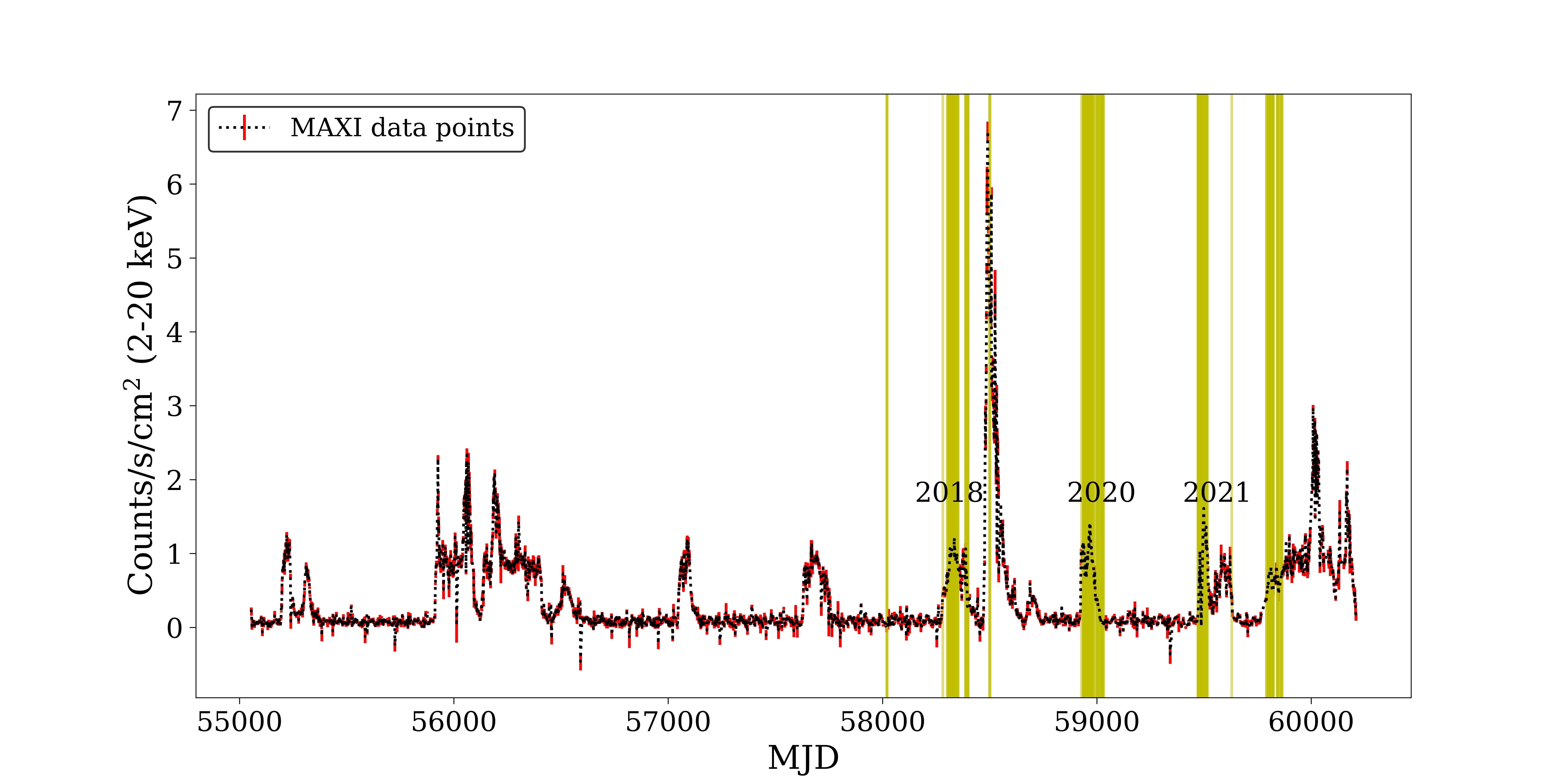}
  \caption[width=\textwidth]{Long-term \emph{MAXI} light curve (black) depicting the flux variations of 4U 1630–472 in the 2.0--20.0 keV energy range. The vertical lines (yellow) indicate the precise dates of the observations conducted by the \emph{NICER} instrument.}
  \label{fig1}
\end{figure*}

The X-ray Timing Instrument (XTI) onboard \emph{NICER} features an array of 56 X-ray ``concentrator'' optics (XRC) matched with silicon drift detectors (SDD). Each XRC efficiently collects X-rays from a significant $30' \times 30'$ field of the sky, focusing them onto an SDD \citep{Okajima}. These detectors excel at recording individual photons, offering excellent spectral resolution and unparalleled time accuracy, with detections precise to 100 ns \citep{gregory}, facilitating the detailed study of rapid X-ray variability during the BHXB outbursts. Additionally, the energy resolution of \emph{NICER} within the 0.2--12 keV energy range proves to be highly advantageous for investigating the energy dependence and spectral properties of the X-ray emission \citep{Gendreau}.

We extracted the \emph{MAXI} light curve as shown in Figure~\ref{fig1}, covering the 2.0--20.0 keV energy range and spanning observations from 2009 onwards, to investigate the long-term behaviour of the black hole source \src. By comparing \emph{NICER} observation dates (depicted by the yellow vertical lines in Figure~\ref{fig1}) against the \emph{MAXI} light curve, we identified three notable outbursts occurring in 2018, 2020, and 2021, with simultaneous \emph{NICER} coverage.
During the 2018 outburst, \emph{NICER} first observed the source on June 11, 2018, and subsequently conducted 47 observations until October 8, 2018. For the 2020 outburst, \emph{NICER} performed 59 observations from March 19 to July 1, 2020. In the year 2021, \emph{NICER} observed the source during its outburst on 30 occasions from September 13 to October 28. A total of 136 observations were obtained across the combined three outbursts. However, for this work, we applied a rigorous criterion where only observations with a total exposure time exceeding 1000 s were considered. This criterion ensured a superior signal-to-noise ratio (SNR) for the resultant power spectra, enhancing the reliability and robustness of our subsequent timing analysis. Following this selection condition, we were left with almost 56 \% of the observations taken by \emph{NICER}, out of which 28 observations belonged to the 2018 outburst, 29 to the 2020 outburst, and 19 observations to the 2021 outburst. Moreover, \emph{NICER} also observed the bright outburst in 2019 for three days during its peak, however, since the outbursts and the quasi-periodic variabilities in 2018, 2020, and 2021 had not yet been studied using \emph{NICER} observations, we decided to focus on these three outbursts for our analysis. We did not detect any QPOs in the three observations of the 2019 outburst, thus making the 2019 outburst irrelevant, given the focus of this paper.

For the timing analysis, we used the high-time resolution event mode data from the XTI onboard \emph{NICER}. The Level 1 data was processed through the \emph{NICER} Level2 pipeline, specifically the {\tt NICERL2}\footnote{\href{https://heasarc.gsfc.nasa.gov/docs/nicer/analysis threads/nicerl2/}{https://heasarc.gsfc.nasa.gov/docs/nicer/analysis threads/nicerl2/}} task, to obtain the clean event files. This processing ensured the removal of instrumental artifacts and certain background noise, enabling reliable timing analysis of the variability of the source. For the temporal analysis, we considered the full energy range to maximize signal-to-noise ratio (SNR). Barycentre correction was performed for all the level 2 filtered files using the task {\tt BARYCORR}. We applied further GTI (good time interval) filtering on our dataset, by eliminating the data gaps and instrument noise times, and subsequently extracted the \emph{NICER} long-term light curves for each individual outburst as shown in Figures \ref{lc2018}, \ref{lc2020} and \ref{lc2021}.

For the spectral analysis,  we utilized the advanced spectral capabilities of the \emph{NICER} instrument to investigate the radiative behaviour of the black hole source \src during its outbursts in 2018, 2020, and 2021. To generate the required spectra, as well as the energy-channel response (rmf) and ancillary response (arf) files, we employed {\tt HEASOFT version 6.31.1}\footnote{\href{https://heasarc.gsfc.nasa.gov/docs/software/lheasoft/}{https://heasarc.gsfc.nasa.gov/lheasoft/download.html}} and {\tt CALDB version 20221001}
\footnote{\href{https://heasarc.gsfc.nasa.gov/docs/heasarc/caldb/nicer/}{https://heasarc.gsfc.nasa.gov/docs/heasarc/caldb/nicer/}}.
To estimate the background spectra, we utilized the \emph{NICER} background model 3C50 \citep{Remillard}. The spectral parameters were determined by fitting the spectra using {\tt XSPEC version 12.13.0c} \footnote{\href{https://heasarc.gsfc.nasa.gov/xanadu/xspec/issues/issues.html}{https://heasarc.gsfc.nasa.gov/xanadu/xspec/issues/issues.html}}. The {\tt NICERL3-SPECT}\footnotemark[4]\textsuperscript{,}\footnote{\href{https://heasarc.gsfc.nasa.gov/docs/nicer/analysis_threads/nicerl3-spect/}{https://heasarc.gsfc.nasa.gov/docs/nicer/analysis\_threads/nicerl3-spect/}} task within the {\tt NICERL3} pipeline was employed to generate source and background spectra, along with responses, in the 0.5--12.0 keV energy range. For the spectral analysis, we used the 2-10 keV band to avoid observed systematics and instrumental residuals below 2 keV and above 10 keV, which allowed for improved fitting reliability\footnote{\href{https://heasarc.gsfc.nasa.gov/docs/nicer/analysis_threads/arf-rmf/}{https://heasarc.gsfc.nasa.gov/docs/nicer/analysis\_threads/arf-rmf/}}\textsuperscript{,}\footnote{\href{https://heasarc.gsfc.nasa.gov/docs/nicer/data_analysis/workshops/2024/joint2024.html}{Craig Markwardt's slides from the Joint NICER/IXPE Workshop 2024}}. For generating spectra for each orbit in any particular observation, when required for the time-resolved spectral analysis, we generated the spectra using {\tt XSELECT} on the level2 event files. Since \texttt{NICERL3-SPECT} automatically applies \emph{NICER}-recommended systematic errors to the data, we additionally introduced a conservative systematic error of 1.5\%\footnote{\href{https://heasarc.gsfc.nasa.gov/docs/nicer/analysis_threads/cal-recommend/}{NICER Calibration Recommendations}} to the spectra obtained using \texttt{XSELECT}. Furthermore, the spectral data from {\tt XSELECT} and \textbf{{\tt NICERL3-SPECT}} involved grouping the original channels such that each new bin contained at least 10 counts to ensure sufficient SNR per energy bin. This approach ensured that the spectra generated by both {\tt NICERL3-SPECT} and {\tt XSELECT} were treated consistently, thus, allowing for reliable comparisons and unbiased analysis. In this work, the updated version of {\tt NICERL3-SPECT} (as of February 6th, 2023) was confirmed to produce consistent results with the aforementioned spectral generation technique.

%% file: results.tex
\section{ANALYSIS AND RESULTS}\label{sec:results}
\subsection{Fundamental diagrams}

\begin{figure*}[h]
  \centering
  \begin{minipage}[t]{0.45\textwidth}
    \centering
    \begin{subfigure}[t]{\linewidth}
      \includegraphics[width=\linewidth]{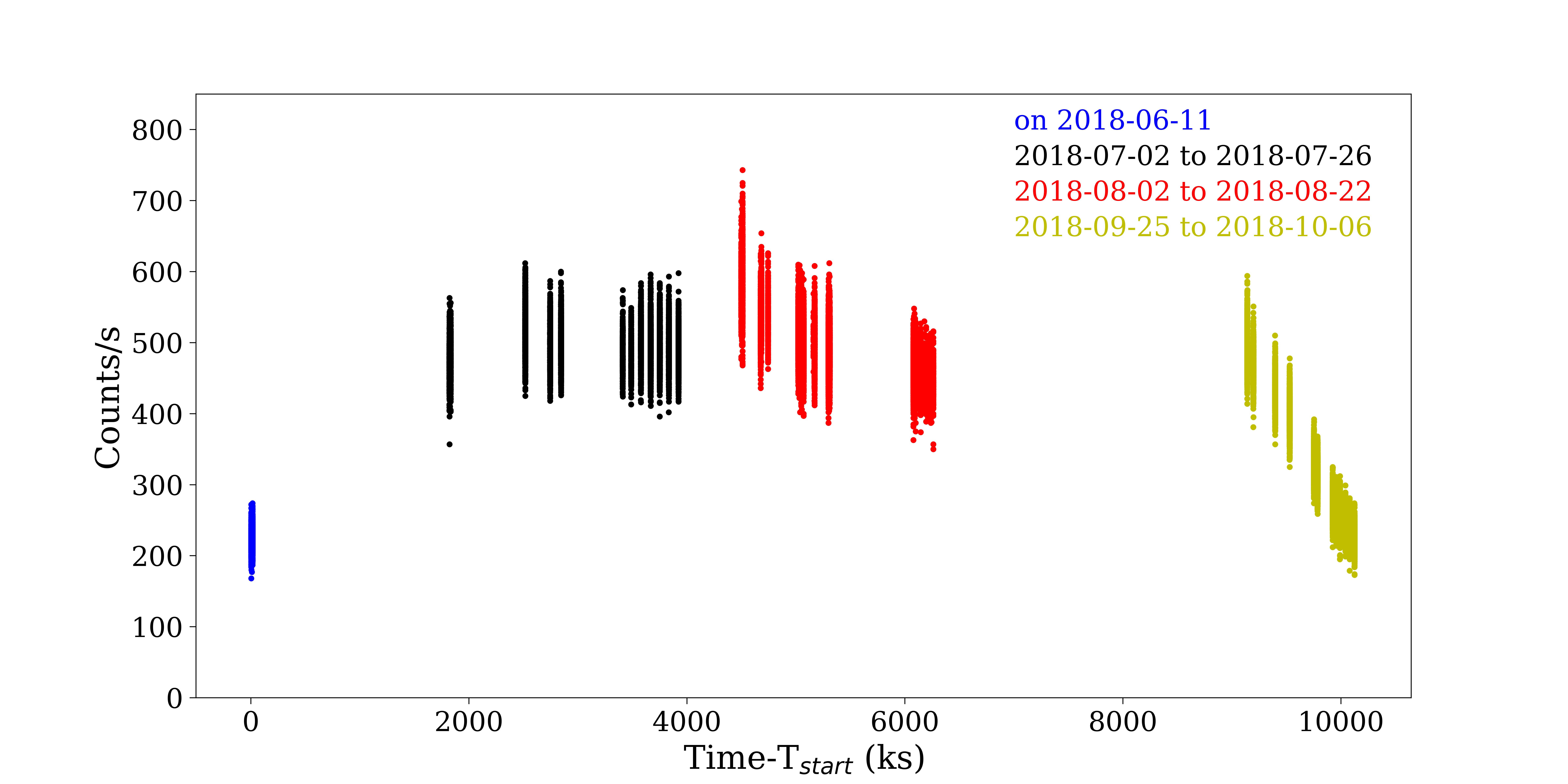}
      \caption{2018}
      \label{lc2018}
    \end{subfigure}
    \vspace{\baselineskip} 
    \begin{subfigure}[t]{\linewidth}
      \includegraphics[width=\linewidth]{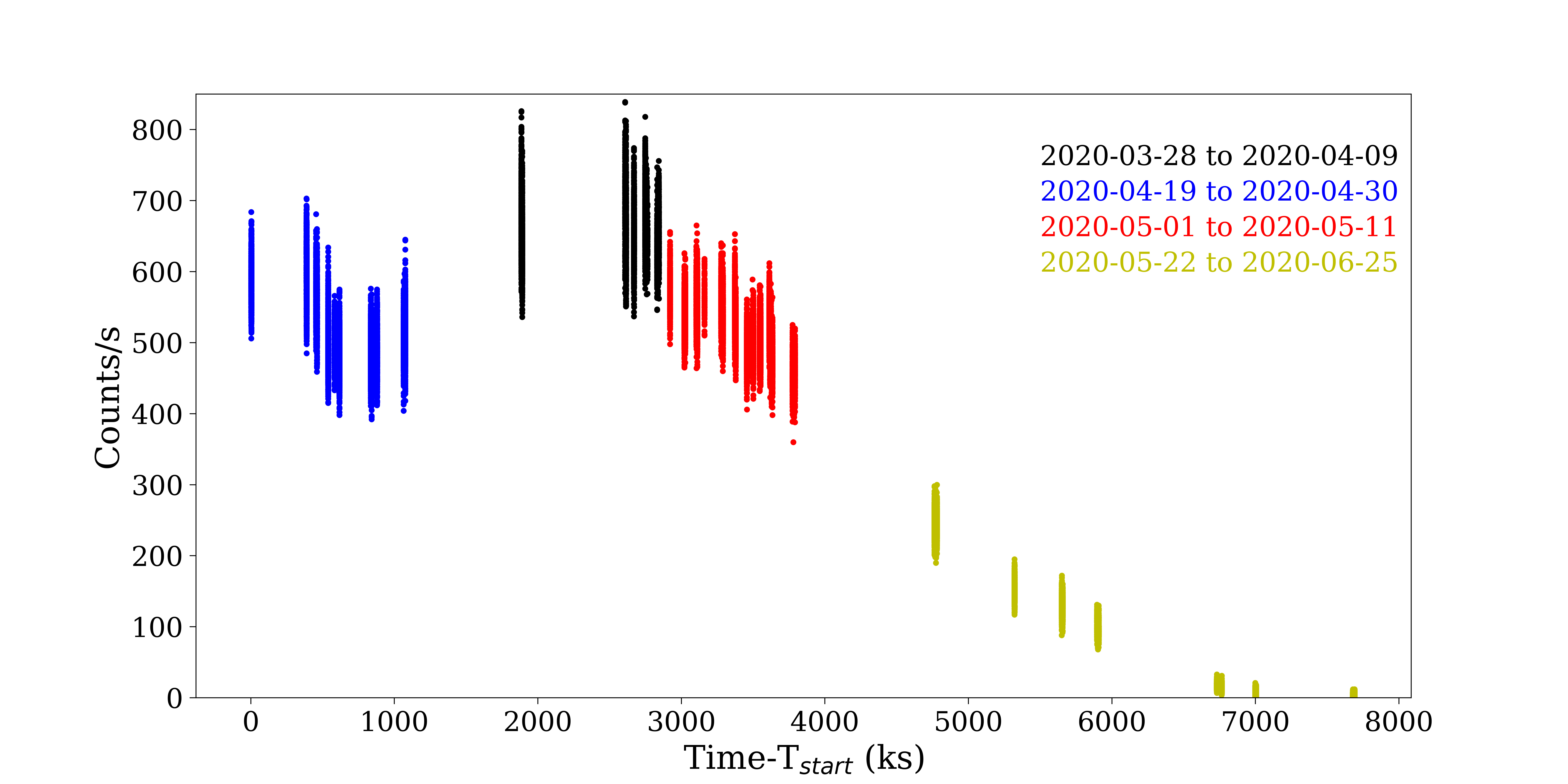}
      \caption{2020}
      \label{lc2020}
    \end{subfigure}
    \vspace{\baselineskip} 
    \begin{subfigure}[t]{\linewidth}
      \includegraphics[width=\linewidth]{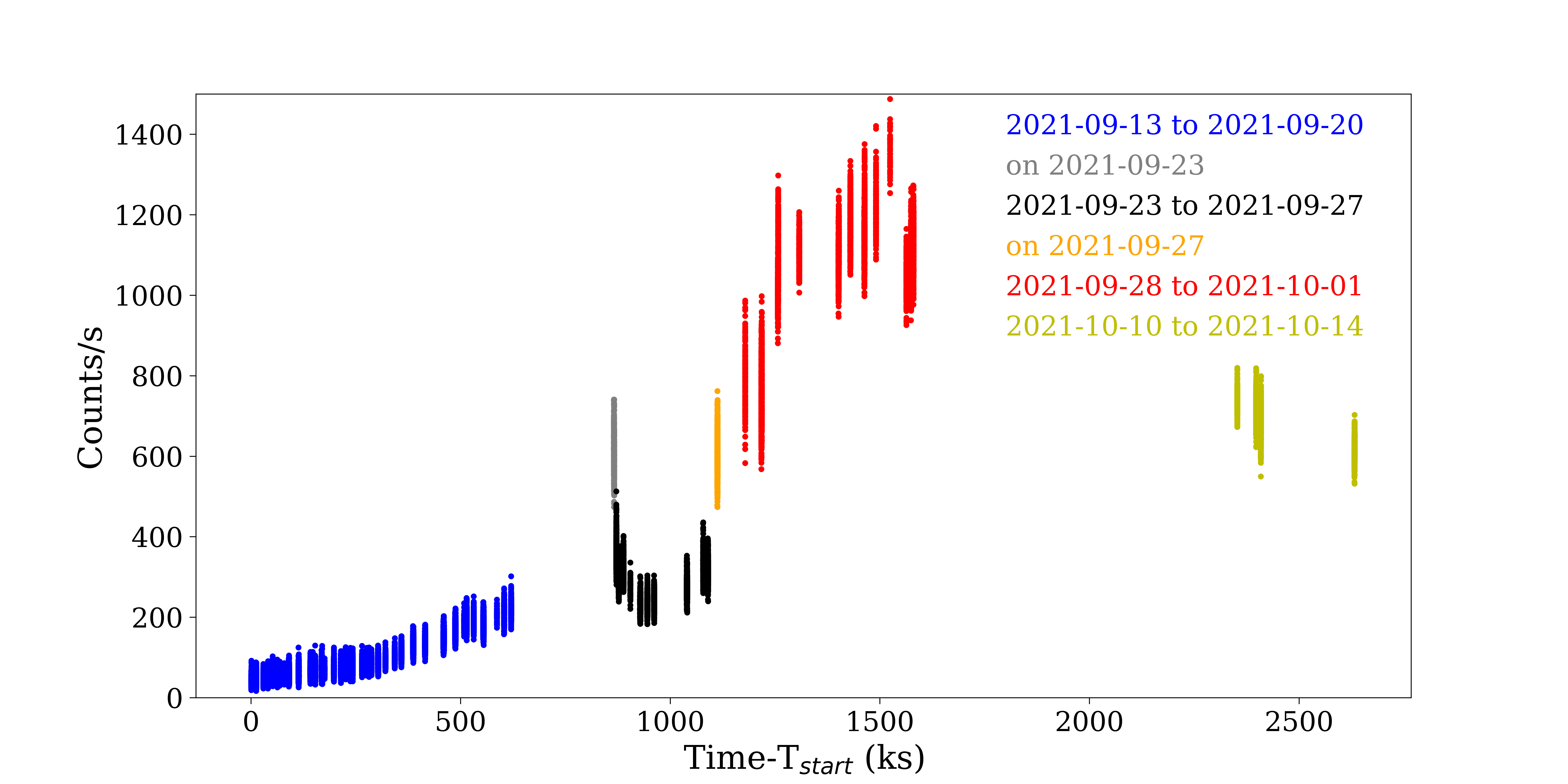}
      \caption{2021}
      \label{lc2021}
    \end{subfigure}
  \end{minipage}%
  \hfill
  \begin{minipage}[t]{0.45\textwidth}
    \centering
    \begin{subfigure}[t]{\linewidth}
      \includegraphics[width=\linewidth]{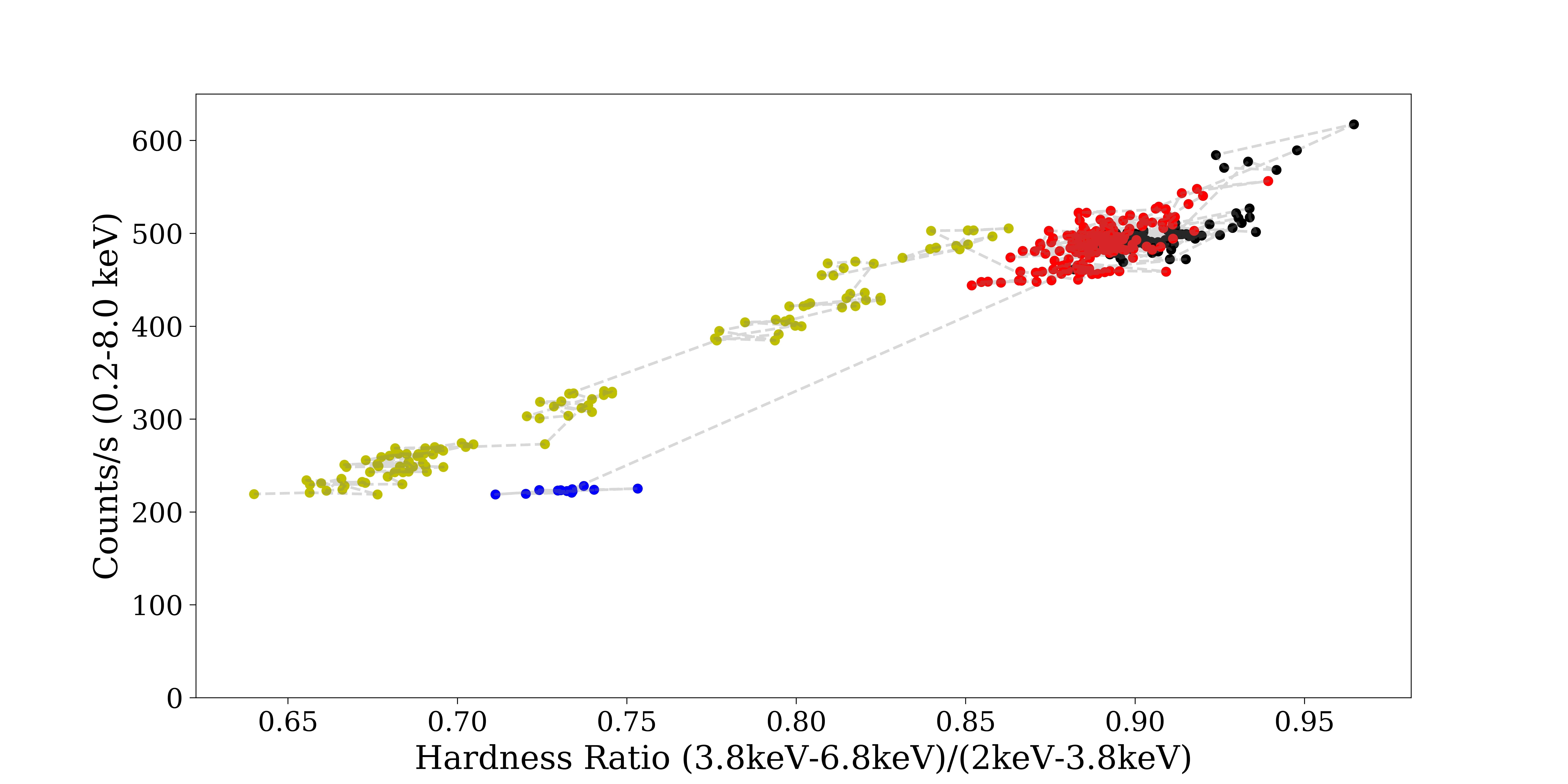}
      \caption{2018}
      \label{hid2018}
    \end{subfigure}
    \vspace{\baselineskip} 
    \begin{subfigure}[t]{\linewidth}
      \includegraphics[width=\linewidth]{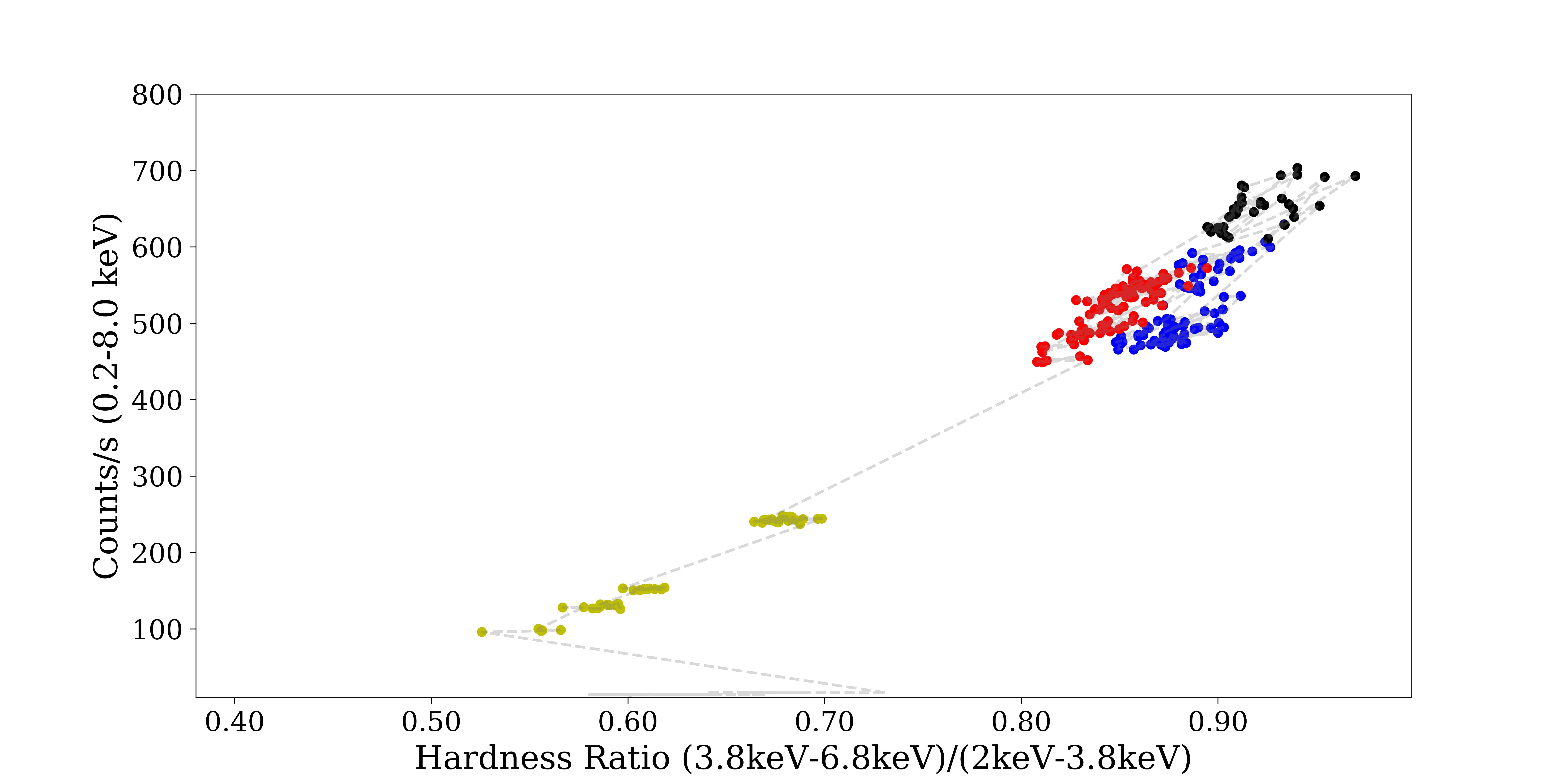}
      \caption{2020}
      \label{hid2020}
    \end{subfigure}
    \vspace{\baselineskip} 
    \begin{subfigure}[t]{\linewidth}
      \includegraphics[width=\linewidth]{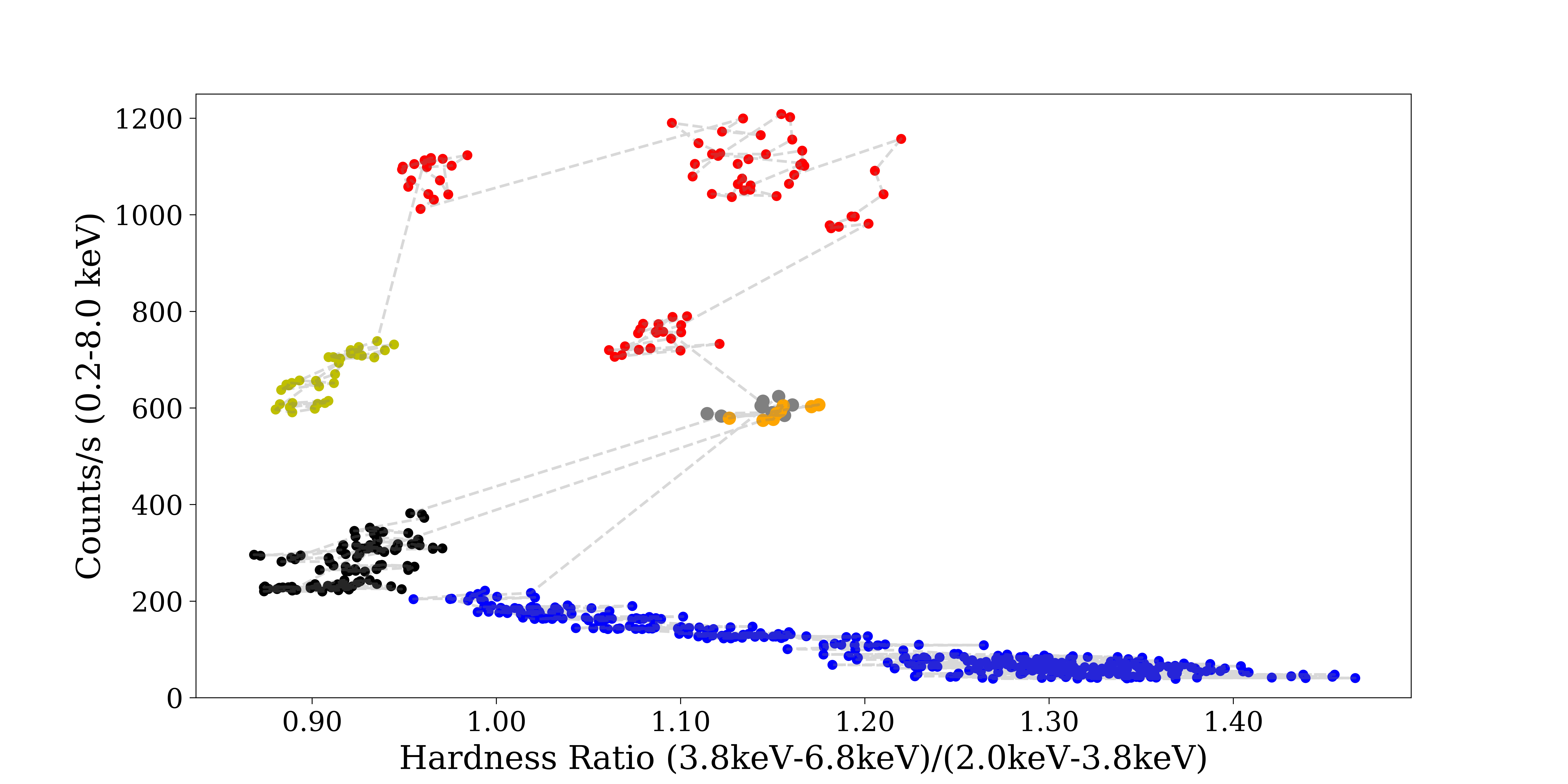}
      \caption{2021}
      \label{hid2021}
    \end{subfigure}
  \end{minipage}
  \caption{Light curves (left) and HIDs (right) of the 2018, 2020, and 2021 outbursts of \src observed by \emph{NICER}. The color code used for HIDs and light curves represents different segments corresponding to distinct observation dates and is the same for each outburst, ensuring consistency and facilitating a direct comparison between the light curves and the associated hardness ratios. The gray and orange points in (c) and (f) refer to the observations corresponding to MJD 59480.793 and 59483.648, respectively.}
\end{figure*}

The light curves extracted from the \emph{NICER} event data for the three outbursts considered for this study are depicted in Figures \ref{lc2018}, \ref{lc2020} and \ref{lc2021}, where the different colors represent certain distinct observational epochs during the outburst where the intensity showed a consistent behaviour. In Figure~\ref{lc2018}, \emph{NICER} primarily captured the high-flux state of the 2018 outburst and the subsequent decay phase of the secondary outburst. Figure \ref{lc2020} illustrates the transitional phase of the source from the primary to the secondary outburst during the 2020 outburst event, accompanied by its subsequent decay phase. The 2021 outburst, as shown in Figure \ref{lc2021}, was predominantly observed during its rising phase. Detailed investigation of individual observations providing a comprehensive understanding of the spectro-temporal evolution of the emission characteristics of the source throughout the outbursts can be found in sections $\S$\ref{result: critical frequency} and $\S$\ref{result: flare and transition}.

For each of the three outbursts depicted in Figures~\ref{lc2018}, \ref{lc2020} and \ref{lc2021}, we constructed the Hardness Intensity Diagrams (HIDs) based on the level 2 \emph{NICER} data, as illustrated in Figures~\ref{hid2018}, \ref{hid2020} and \ref{hid2021} respectively. The HIDs provide a preliminary view of the spectral evolution and intensity variations of the source during the outbursts. The Hardness Ratio (HR) was calculated using the ratio of intensities in the soft X-ray energy band of 2 keV to 3.8 keV and in the hard energy band of 3.8 keV to 6.8 keV. The average count rate, obtained from the energy range of 0.2 keV to 8 keV, was used to determine the intensity. Each point in the HID represents the HR and count rate of the light curve, with a bin size of 128 s. 

To incorporate the chronological information of the spectral and the intensity evolution, a color scheme was adopted as a third axis, representing the different observation times. This color scheme aligns with the chronological color scheme employed in the light curves (Figures~\ref{lc2018}, \ref{lc2020} and \ref{lc2021}), facilitating the tracking of the spectral evolution of the source throughout the three outbursts.

\begin{figure}
  \centering
  \includegraphics[width=\linewidth]{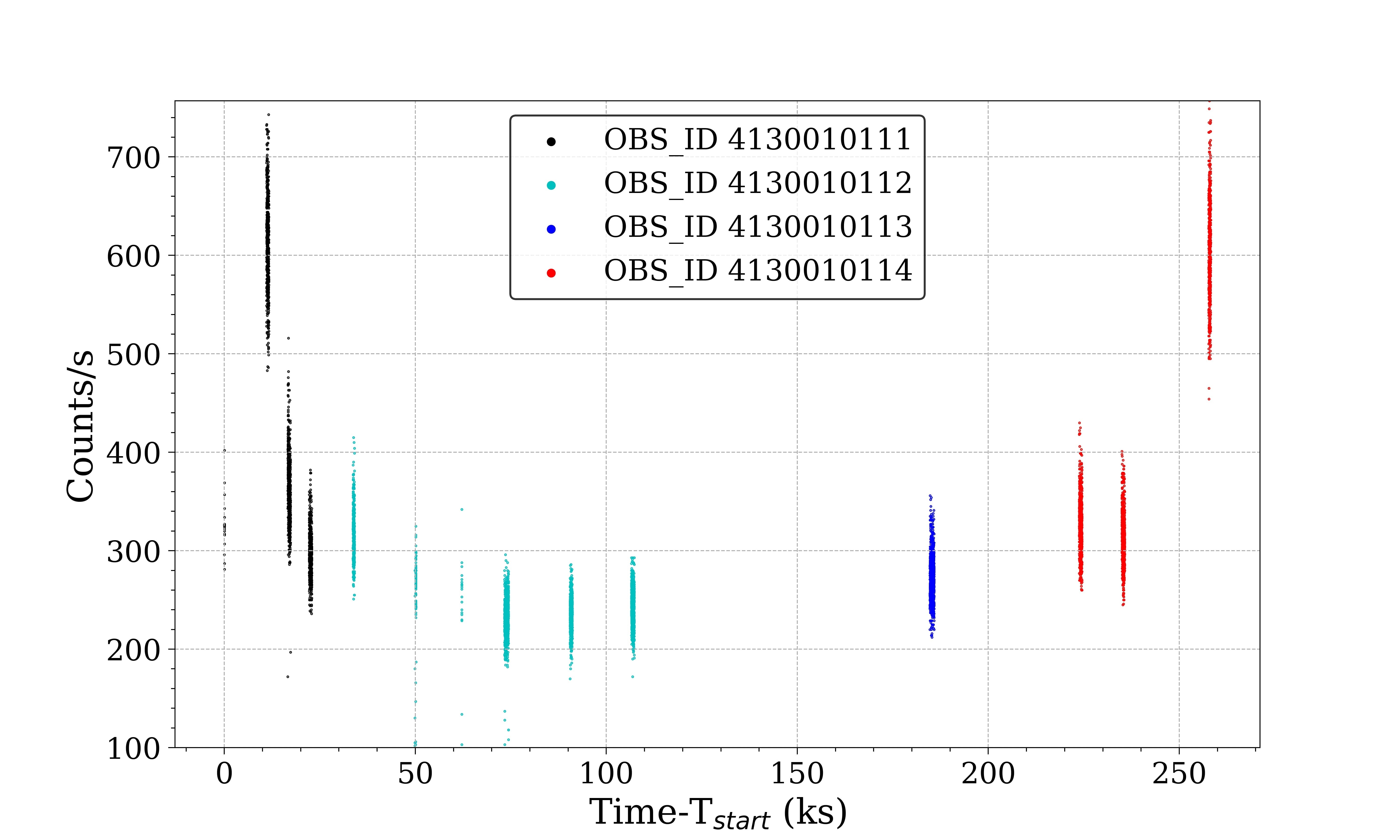}
  \centering
  \caption{The light curve corresponding to four consecutive observations from the 2021 outburst (from the gray segment to the orange segment in Figure~\ref{lc2021}) revealing significant fluctuations in the count rate over short time scales. Notably, the count rate exhibited an approximately twofold change during the initial three segments of observation 4130010111 (illustrated in black) and the final two segments of observation 4130010114 (depicted in red). }

  \label{LC123-126}
\end{figure}

Notably, within the 2021 outburst, two observations display significant changes in count rates within a single day, as denoted by gray and orange points at times around 800 ks and 1200 ks, respectively in Figure~\ref{lc2021}. Figure \ref{LC123-126} shows the variation of count rates with time, where the different observation IDs falling within these two observations are depicted by different colors. The effect of the two observations with sudden flux changes can be seen in the evolution observed in the HID (see gray and orange circles in Figure~\ref{hid2021}). The HID displayed notable sharp transitions, indicating significant changes in the spectral behaviour of the source on a timescale of a few hours. The consequences of the transitions in these observations on the spectral and timing parameters are discussed in detail in sections \S\ref{result: flare and transition} and \S\ref{discussion: flares and transitions}.

\subsection{Timing Analysis}

We examined the X-ray variability characteristics of the emission of the dynamic BHXB \src through an extensive temporal analysis.  For every observation, the power density spectra (PDS) was obtained from every 64 s of the light curve applying a time resolution of 1/64 s, and then the resultant PDS were averaged to obtain the time-averaged PDS. The time-averaged PDS warrants a reduced noise contribution. Frequency averaging \citep{klis1988} was also applied to increase further the SNR, especially in the case of QPOs observed at a frequency higher than 2 Hz. Leahy-normalization was implemented on the resultant power spectra to ensure a known analytical representation of the Poissonian noise power distribution \citep{leahy1983, klis1988}.
Following this method, the PDS was seen to exhibit distinct variability features, including broadband components, peaked noise components, and Poisson noise. The Poisson noise power, consistent with expectations for a Leahy-normalized PDS, approached a value of 2 and was fitted with a constant model. The broadband noise was well-fitted using either a power law or a zero-centered Lorentzian. The peaked noise components were described by Lorentzian as described in \cite{belloni2002}, characterized by parameters such as the total RMS amplitude ($r$), the FWHM (full width at half maximum) ($\Delta$), and the centroid frequency ($\nu_{0}$). The $\chi^{2}$ minimization method was employed to find the best-fit model for the PDS. The uncertainties for these parameters were determined at a $1~\sigma$ confidence level.
The Q-factor, represented by \(\nu_0/2\Delta\), serves as a measure of the coherence of the signal in the PDS \citep{belloni&tomaso}. For the Q-value, if the lower limit (i.e., $Q - \delta Q$) for a peaked noise component exceeded 2, it was classified as a QPO.\\
\begin{figure}
    \centering
    \includegraphics[width=\linewidth]{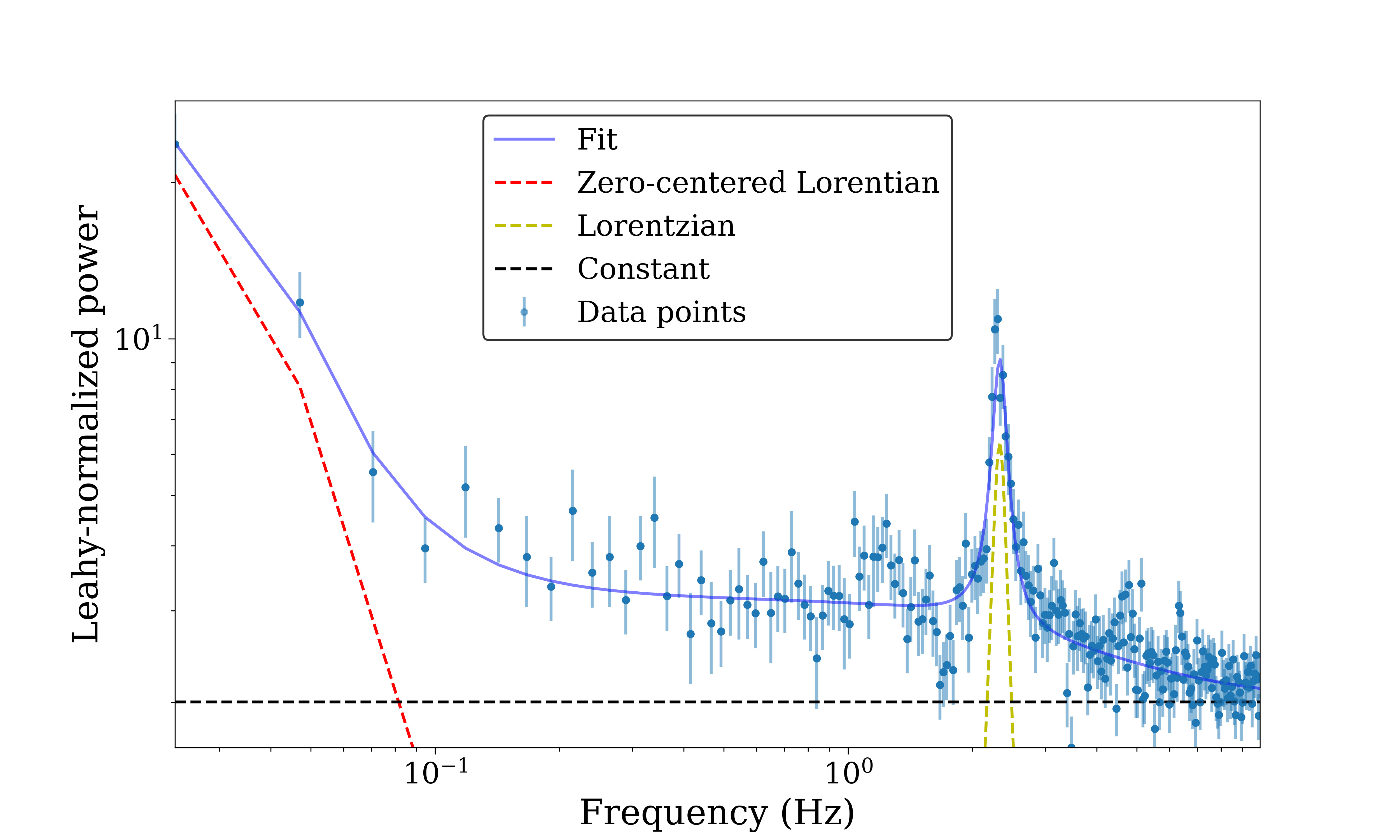}
    \caption{The power density spectra (PDS) of the type-C QPO observed on MJD 59476.407. The continuum is fitted with a zero-centered Lorentzian (in red) along with a constant (in black) to account for the Poisson noise. A Lorentzian (in yellow) is used to fit the QPO centred at $\sim 2.31$ Hz. The combined best-fit model (reduced $\chi^2=1.14$) is depicted by the solid blue line.}
    \label{pdsfitting}
\end{figure}

\subsubsection{Energy-integrated PDS} \label{result: energy averaged pds}
We first constructed the energy-integrated PDS for each observation, considering the full \emph{NICER} energy range of 0.2--12.0 keV. Figure~\ref{pdsfitting} shows a representative example of the PDS modelling for the observation taken on MJD 59476.407, where the continuum variability is fitted with a zero-centered Lorentzian and a constant, and the QPO is fitted with a Lorentzian \citep{belloni2002}. The PDS has been logarithmically binned, and the frequency ranges of Figure~\ref{pdsfitting} have been selected to highlight the region of interest of the PDS, especially the QPO. Following this methodology, we obtained single QPO detection in the SIMS and the HSS during the outbursts of 2018 and 2020, respectively. 19 QPOs were initially detected during the 2021 outburst, out of which 11 were confirmed as type-C QPOs observed in the LHS and HIMS. During the rising phase of the 2021 outburst, the {\em NICER} instrument captured the transition of the source from the LHS to the HSS. Initially, we extracted the PDS from all single-day observations during the outburst using \emph{NICER}. Notably, among the 11 type-C QPO observation IDs, 7 single-day observations initially displayed the presence of multiple QPOs that were very closely spaced in frequency, with none exhibiting harmonics.  

\begin{figure}
  \centering
  \includegraphics[width=\linewidth]{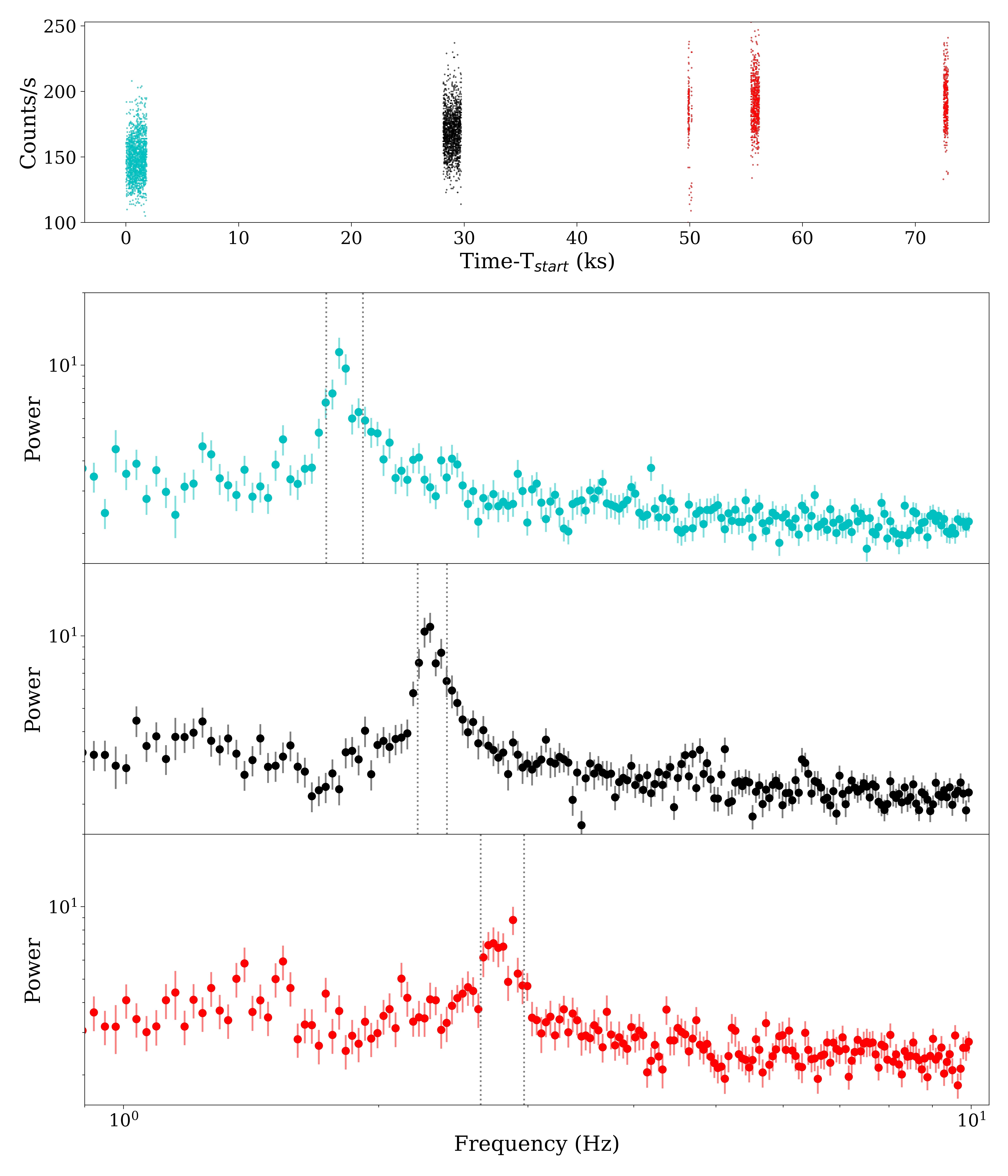}
  \caption{The light curve of the 2021 outburst from the \emph{NICER} X-ray observation made on MJD 59476 (OBSID 4130010107) is presented in the upper panel. The light curve is divided into different orbit segments, each represented by a different color. The bottom three panels display the corresponding power spectra for each of these segments, represented by the same color code. Notably, the power spectra reveal a remarkable phenomenon wherein the average count rate, as well as the central frequency of the QPO, exhibits variations even within a single observation. See Table \ref{table1} for the timing parameters of these three orbit observations.}
  \label{OBSID4130010107}
\end{figure}

\textbf{\textcolor{black}{Rapid evolution of QPOs in the 2021 outburst:}}
To gain deeper insights into this QPO behaviour, we extracted individual PDS from the light curves of each orbit within an observation. Such an investigation has the potential to unveil any dynamic behaviour of the QPO characteristics over short timescales within an observation. 
Through the detailed time-resolved analysis, we observed that the cases with multiple QPO detections in the time-integrated PDS can be explained by a single QPO evolving in frequency throughout the entire observation over the different orbits.
An illustrative example can be observed in Figure \ref{OBSID4130010107}. The light curve displayed in the top panel of Figure \ref{OBSID4130010107} corresponds to the observation taken on MJD 59476 (also see Table~\ref{table1}) where the count rate exhibits a variable behaviour, progressively increasing from one orbit to the other. Figure \ref{OBSID4130010107} also depicts the PDS of the individual orbit segments (different orbits identified with different colors), revealing the presence of a QPO that evolves within the observation (bottom panel). This time-resolved analysis further demonstrated the sensitivity of the QPO frequency to the count rate, allowing us to meticulously track the evolution of this single QPO within each segment. 

Although the time-resolved analysis demonstrated variability within each observation, obtaining PDS from each orbit resulted in a decrease in the SNR of the detection. However, it proved advantageous in tracing the frequency evolution of the QPO. Consequently, we successfully identified a total of 21 type-C QPOs during the LHS and HIMS state of the 2021 outburst of the source \src. To assess the strength of the QPO, the fractional RMS of the QPO was computed. The fractional RMS was background corrected following the method from \cite{bu2015}. The detection significance for each QPO is given by the ratio of the integral power of the QPO to its 1$\sigma$ error  \citep{motta2011}, where the integral power of the QPO was obtained by computing the square of the fractional RMS ($r$) of the QPO.
For the 2021 outburst, we even extracted the PDS for observations that did not have exposure greater than 1000 s, but did not detect any additional QPOs. The temporal parameters and significance related to the QPO fitting and the corresponding average intensity are presented in Table~\ref{table1}.

\begin{figure}
  \centering
  \includegraphics[width=\linewidth]{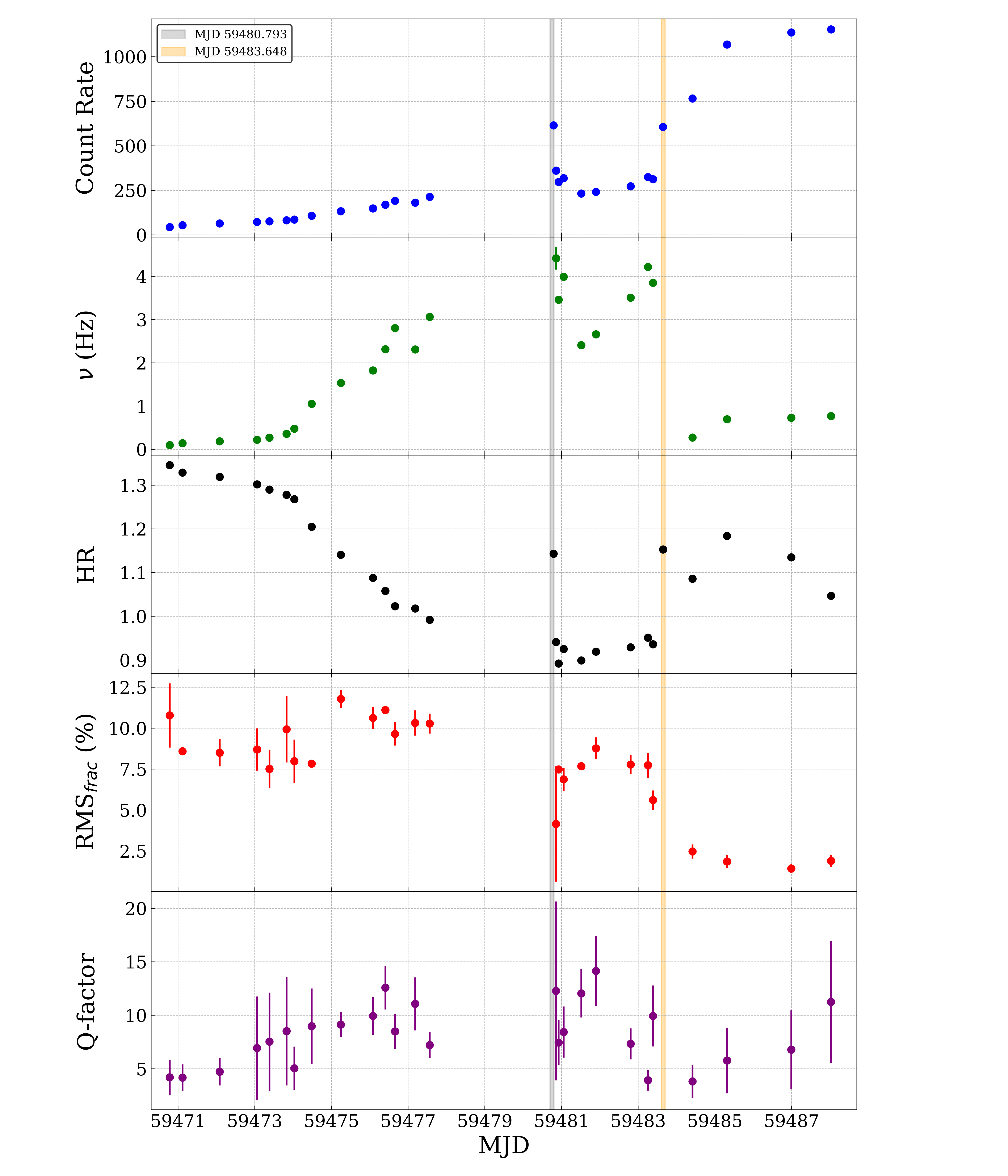}
  \centering
  \caption{\small  
  Temporal variations of key timing parameters of the QPOs detected during the 2021 outburst.
  From top to bottom: Count Rate (0.2--12 keV), QPO centroid frequency, Hardness Ratio (3.8--6.8 keV/2.0--3.8 keV), Fractional RMS, and Q-factor are plotted with time. Regions marked in gray and orange shades indicate dates when the Count Rate increases along with the HR, leading to the disappearance of the QPO.} 
  \label{qpo_parameters}
\end{figure}

\begin{figure}
  \centering
  \includegraphics[width=\linewidth]{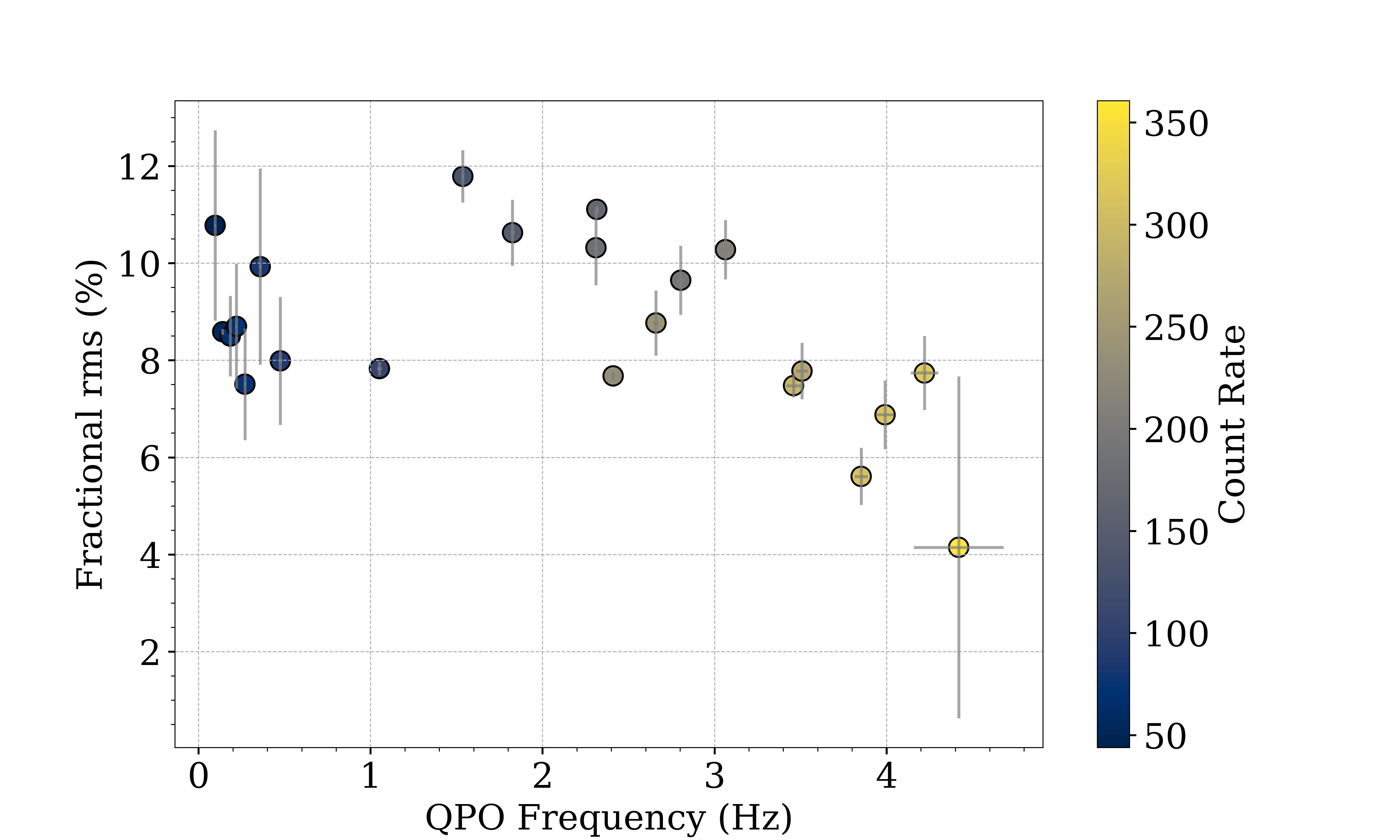}
  \centering
  \caption{\small  QPO fractional RMS as a function of the centroid frequency for the type-C QPOs observed in 2021. The shade of the data points indicates the mean count rate, as elaborated by the color bar.} 
  \label{qpoVSRMSVScr}
\end{figure}

Figure \ref{qpo_parameters} provides insights into the temporal behavior of these timing parameters, offering a deeper understanding of the observed evolution of the QPOs during the 2021 outburst. Starting from the top, we observe the count rate, represented by the blue points, estimated considering the overall energy range of 0.2--12 keV. The green points depict the QPO frequency, while the black points represent the Hardness Ratio, indicating the relative intensities in the hard X-ray band (3.8--6.8 keV) and the soft X-ray band (2.0--3.8 keV). The red points represent the fractional RMS, reflecting the amplitude of the QPO. Lastly, the purple points represent the Q-factor, derived from the ratio of the QPO frequency to its FWHM. Figure \ref{qpoVSRMSVScr} displays the relationship between the QPO fractional RMS amplitude and its frequency. The color of the data points corresponds to the average count rate, providing valuable insights into the influence of the overall source intensity on the QPO strength.

\subsubsection{Energy-resolved PDS} \label{result: energy resolved pds}
To understand the behavior of X-ray flux variations at different X-ray energies, we investigated the energy-resolved PDS of the QPOs observed in the rising phase of the 2021 outburst.
To achieve this, we selected three distinct energy bands of the {\em NICER} instrument: the Low Energy ({\bf LE}) band ranging from 0.5--3 keV, the Middle Energy ({\bf ME}) band ranging from 3--6 keV, and the High Energy ({\bf HE}) band ranging from 6--10 keV. The choice of these energy bands allows us to investigate different physical processes contributing to the X-ray emission. 
The LE band primarily captures the thermal emission originating from the accretion disk surrounding the black hole. Conversely, the HE band predominantly represents the inverse-Comptonized emission from the corona. The ME band, occupying an intermediate energy range, may exhibit contributions from both the accretion disk and the corona, making it crucial for analysing the variability properties in \src.

\begin{figure}[h]
  \centering
  \includegraphics[width=1.0\linewidth]{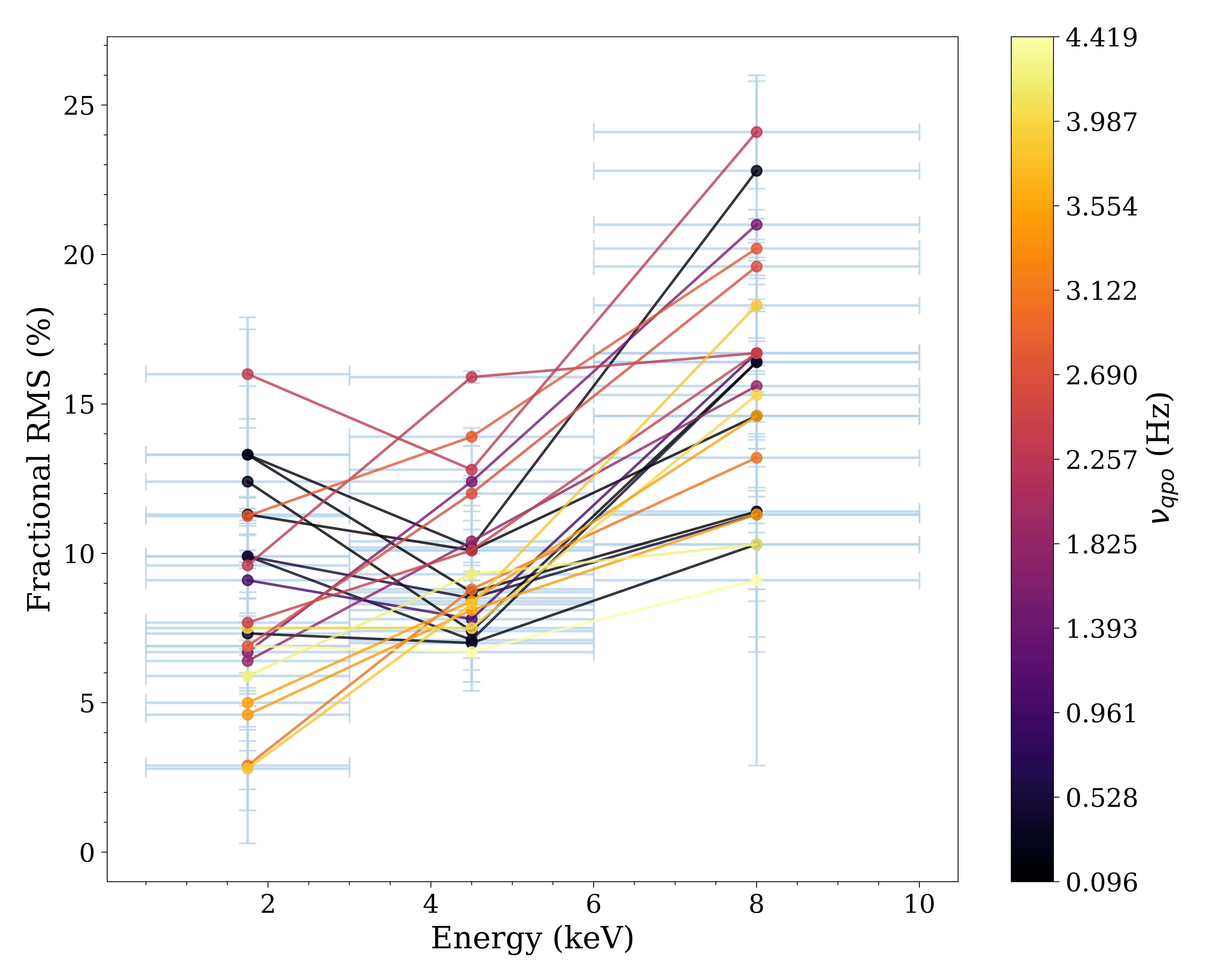}
  \centering
  \caption{\small RMS spectra showing the fractional RMS (in \%) of all the QPOs detected during the rising phase of the 2021 outburst, across the three {\em NICER} energy bands 0.5--3 keV, 3--6 keV and 6--10 keV. The color in the plot represents the centroid frequency of the QPOs, following the scale depicted in the color bar on the right.} 
  \label{erps}
\end{figure}

We computed the PDS in the specified energy ranges and determined the RMS of the QPOs. The construction of the PDS and subsequent estimation of fractional RMS of QPOs in the individual energy bands followed a similar approach as mentioned in $\S$\ref{result: energy averaged pds}.
Figure \ref{erps} presents the RMS spectra of all the QPOs detected during the 2021 outburst, with the color representing the QPO frequency. We observed fractional RMS values exceeding 10\% in the HE band, which is a characteristic of type-C QPOs. 
Notably, for the majority of observations, the fractional RMS of the QPO increased with photon energy, indicating a potential association between the QPO origin and the emission in the 6--10 keV energy range. 
To gain further insights, we performed a time lag analysis on the QPO observations, focusing on the three specific energy bands (0.5--3 keV, 3--6 keV, and 6--10 keV), also used for energy-resolved power spectral analysis. However, the estimated time lags exhibited significant uncertainties, with most values consistent with zero considering $1~\sigma$ uncertainty.

\begin{figure*}
  \centering
  \begin{threeparttable}
    \renewcommand{\arraystretch}{1.5} 
    \captionof{table}{Parameters for QPOs observed over observations of the different outbursts}
    \footnotesize
    \begin{tabular*}{\textwidth}{@{\extracolsep{\fill}}|c|c|c|c|c|c|c|}
    \hline
    \textbf{Observation ID} & \textbf{MJD} & \textbf{Mean CR} & \textbf{$\boldsymbol{\nu_{qpo}}$ (Hz)}\tnote{a} & \textbf{Q-Factor}\tnote{a} & \textbf{RMS$_{\boldsymbol{frac}}$ (\%)}\tnote{a} & \textbf{Significance ($\boldsymbol{\sigma}$)}\tnote{b}\\
    \hline
    \multicolumn{7}{c}{\textbf{2018 Outburst} \tnote{a}} \\
    \hline
    \multirow{1}{*}{1130010124} & 58325.930 & 492.16 $\pm$ 1.32 & 0.252 $\pm$ 0.007 & 3.503 $\pm$ 1.352 & 2.91 $\pm$ 0.30 & 4.85\\
    \hline
    \multicolumn{7}{c}{\textbf{2020 Outburst}}\\
    \hline
    \multirow{1}{*}{3130010128}& 58966.566 & 669.03 $\pm$ 1.78 & 0.124 $\pm$ 0.003 & 6.192 $\pm$ 3.289 & 8.68 $\pm$ 1.16 & 3.74\\
    \hline   
    \multicolumn{7}{c}{\textbf{2021 Outburst}} \\
    \hline
    \multirow{1}{*}{4130010101} & 59470.783 & 43.93 $\pm$ 0.31 & 0.096 $\pm$ 0.004 & 4.201 $\pm$ 1.647 & 10.78 $\pm$ 1.96 & 2.75 \tnote{c}\\
    \hline
    \multirow{1}{*}{4130010102} & 59471.115 & 54.25 $\pm$ 0.27 & 0.140 $\pm$ 0.004 & 4.169 $\pm$ 1.253 & 8.59 $\pm$ 0.064 & 67.11\\
    \hline
    \multirow{1}{*}{4130010103} & 59472.085 & 64.61 $\pm$ 0.25 & 0.184 $\pm$ 0.003 & 4.718 $\pm$ 1.272 & 8.50 $\pm$ 0.83 & 5.12 \\
    \hline
    \multirow{3}{*}{4130010104} & 59473.061 & 72.79 $\pm$ 0.56 & 0.220 $\pm$ 0.005 & 6.931 $\pm$ 4.834 & 8.70 $\pm$ 1.29 & 3.37 \\
    & 59473.382 & 75.95 $\pm$ 0.43 & 0.269 $\pm$ 0.004 & 7.535 $\pm$ 4.589 & 7.51 $\pm$ 1.15 & 3.26 \\
    & 59473.830 & 82.35 $\pm$ 0.78 & 0.358 $\pm$ 0.008 & 8.516 $\pm$ 5.069 & 9.93 $\pm$ 2.02 & 2.46 \tnote{c} \\
    \hline
    \multirow{2}{*}{4130010105} & 59474.032 & 86.06 $\pm$ 0.46 & 0.475 $\pm$ 0.011 & 5.041 $\pm$ 2.037 & 7.99 $\pm$ 1.32 & 3.03 \\
    & 59474.485 & 107.34 $\pm$ 0.64 & 1.051 $\pm$ 0.015 & 8.978 $\pm$ 3.529 & 7.83 $\pm$ 0.14 & 27.96 \\
    \hline
    \multirow{1}{*}{4130010106} & 59475.246 & 132.63 $\pm$ 0.51 & 1.536 $\pm$ 0.007 & 9.127 $\pm$ 1.173 & 11.79 $\pm$ 0.54 & 10.91 \\
    \hline
    \multirow{3}{*}{4130010107} & 59476.084 & 148.73 $\pm$ 0.62 & 1.825 $\pm$ 0.012 & 9.946 $\pm$ 1.794 & 10.63 $\pm$ 0.68 & 7.82 \\
    & 59476.407 & 169.08 $\pm$ 0.66 & 2.315 $\pm$ 0.011 & 12.582 $\pm$ 2.037 & 11.11 $\pm$ 0.08 & 69.44 \\
    & 59476.659 & 191.67 $\pm$ 0.84 & 2.752 $\pm$ 0.027 & 5.463 $\pm$ 0.915 & 9.65 $\pm$ 0.71 & 6.79 \\
    \hline
    \multirow{2}{*}{4130010108} & 59477.184 
    & 180.92 $\pm$ 0.83 & 2.310 $\pm$ 0.015 & 11.071 $\pm$ 2.478 & 10.32 $\pm$ 0.77 &  6.70 \\
    & 59477.562 & 213.31 $\pm$ 1.01 & 3.063 $\pm$ 0.023 & 7.210 $\pm$ 1.214 & 10.28 $\pm$ 0.61 & 8.43 \\
    \hline
    \multirow{3}{*}{4130010111} & 59480.793 & 614.74 $\pm$ 2.03 & - & - & - & -  \\ 
    & 59480.857 & 360.61 $\pm$ 1.45 & 4.419 $\pm$ 0.261 & 12.281 $\pm$ 8.362 & 4.15 $\pm$ 3.52 & 0.59 \tnote{c} \\
    & 59480.923 & 296.91 $\pm$ 1.42 & 3.459 $\pm$ 0.042 & 7.452 $\pm$ 2.102 & 7.48 $\pm$ 0.24 & 15.58 \\
    \hline
    \multirow{3}{*}{4130010112} & 59481.055 & 318.35 $\pm$ 1.67 & 3.991 $\pm$ 0.049 & 8.436 $\pm$ 2.388 & 6.88 $\pm$ 0.71 & 4.85 \\
    & 59481.515 & 232.13 $\pm$ 0.77 & 2.410 $\pm$ 0.013 & 12.050 $\pm$ 2.256 & 7.68 $\pm$ 0.09 & 42.66 \\
    & 59481.898 & 241.49 $\pm$ 1.15 & 2.659 $\pm$ 0.016 & 8.436 $\pm$ 2.256 & 8.77 $\pm$ 0.67 & 6.54 \\
    \hline
    \multirow{1}{*}{4130010113} & 59482.803 & 272.96 $\pm$ 0.77 & 3.508 $\pm$ 0.032 & 7.329 $\pm$ 1.451 & 7.78 $\pm$ 0.58 & 6.71 \\
    \hline
    \multirow{3}{*}{4130010114} & 59483.255 & 324.29 $\pm$ 0.99 & 4.220 $\pm$ 0.080 & 3.925 $\pm$ 0.972 & 7.74 $\pm$ 0.76 & 5.09 \\
    & 59483.385 & 312.57 $\pm$ 1.33 & 3.852 $\pm$ 0.038 & 9.939 $\pm$ 2.851 & 5.61 $\pm$ 0.59 & 4.75 \\
    & 59483.648 & 605.87 $\pm$ 2.14 & - & - & - & -  \\
    \hline
    \multirow{1}{*}{4130010115} & 59484.416 & 765.46 $\pm$ 1.48 & 0.271 $\pm$ 0.011 \tnote{d} & 3.821 $\pm$ 1.540 & 2.47 $\pm$ 0.43 & 2.87 \\
    \hline
    \multirow{1}{*}{4130010116} & 59485.316 & 1068.23 $\pm$ 2.11 & 0.695 $\pm$ 0.021 \tnote{d} & 5.774 $\pm$ 3.063 & 1.86 $\pm$ 0.41 & 2.27 \\
    \hline
    \multirow{1}{*}{4130010118} & 59486.992 & 1136.51 $\pm$ 1.58 & 0.729 $\pm$ 0.017 \tnote{d} & 6.786 $\pm$ 3.675 & 1.43 $\pm$ 0.26 & 2.75 \\
    \hline
    \multirow{1}{*}{4130010119} & 59488.028 & 1153.55 $\pm$ 2.06 & 0.767 $\pm$ 0.014 \tnote{d} & 11.248 $\pm$ 5.691 & 1.90 $\pm$ 0.36 & 2.64 \\
    \hline
    \end{tabular*}
    \label{table1}
    \begin{tablenotes}
    \item[a] The uncertainties for these parameters were determined at a $1~\sigma$ confidence level.
    \item[b] The detection significance was calculated by the ratio of the integral power (rms$^{2}$) of the QPO to its 1$\sigma$ error \citep{motta2011}. 
    \item[c] To additionally assess the significance of the fit improvement when introducing an additional Lorentzian for this QPO, an F-test was conducted that gave a chance probability of 4.10$\times$10$^{-3}$, 4.15$\times$10$^{-2}$ and 4.20$\times$10$^{-7}$ for observations taken on MJD 59470.783, 59473.830 and 59480.857 respectively.
    \item[d] The feature was observed during the rising phase, close to the peak of the outburst, but its type is unclear. Detailed discussion is present in $\S$\ref{result: flare and transition}.
    \end{tablenotes}
  \end{threeparttable}%
\end{figure*}

\subsection{Spectral analysis}
To investigate the evolving nature of the QPOs and their link with distinct X-ray emission regions, we conducted spectral analysis of the \emph{NICER} observations during the 2018, 2020, and 2021 outbursts of \src in which QPOs were detected. We restricted our spectral analysis to the 2.0-–10.0 keV energy range, considering the {\tt TBABS*(POWERLAW*DISKBB)} model for the spectral fitting. This model consisted of three components to account for the observed X-ray spectral behavior. The {\tt Tbabs} component represented Galactic absorption due to neutral hydrogen, characterized by the neutral Hydrogen density parameter N$_{H}$, which was kept free in our analysis. The abundance model was set to consider abundances provided in \citet{Wilms}. The thermal emission from the accretion disk was modeled using the multicolor disk blackbody model {\tt DISKBB} \citep{diskbb}, which described the soft X-ray spectral characteristics.
The non-thermal emission from electrons in the corona was incorporated using the {\tt POWERLAW} model, which accounts for the hard X-ray emission resulting from the Comptonization of seed photons. 

In addition, the presence of intrinsic features in the \emph{NICER} spectra between 1.7 keV to 3.5 keV, arising from the detector\footnote{\href{https://heasarc.gsfc.nasa.gov/docs/nicer/analysis_threads/arf-rmf/}{https://heasarc.gsfc.nasa.gov/docs/nicer/analysis\_threads/arf-rmf/}} itself, had an impact on the fitting procedure. These features, including absorption edge features (2.2--3.5 keV) due to gold shell reflectivity and a silicon absorption feature at 1.840 keV \citep{Ludlam}, affected the fit and contributed to the overall $\chi^2$ value to some extent. To estimate the non-thermal and the total X-ray flux, the energy range of 2.0--10.0 keV was employed, and the {\tt cflux} model was utilized.

\begin{figure*}
  \centering
  \begin{threeparttable}
    \renewcommand{\arraystretch}{1.5} 
    \captionof{table}{Spectral parameters of the model {\tt TBABS*(POWERLAW*DISKBB)}for the QPO Observations of the 2018, 2020, and 2021 outbursts}
    \footnotesize 
    \begin{tabular*}{\textwidth}{@{\extracolsep{\fill}}|c|c|c|c|c|c|}
    \hline
    \textbf{Observation ID} & \textbf{MJD} & \textbf{Total Flux ($\boldsymbol{\times 10^{-9} erg/cm^{2}/s})$} & $\boldsymbol{\Gamma}$ & \textbf{Non-Thermal Flux ($\boldsymbol{\times 10^{-9} erg/cm^{2}/s})$} & $\boldsymbol{\chi^{2}~(d.o.f)}$ \\
    \hline
    \multicolumn{6}{c}{\textbf{2018 Outburst} \tnote{a}} \\
    \hline
    \multirow{1}{*}{1130010124} & 58325.930 & $13.18^{+0.55}_{-1.59}$ & $5.661^{+1.442}_{-0.424}$ & $1.88^{+1.90}_{-1.37}$ & 73.73 (103) \\
    \hline
    \multicolumn{6}{c}{\textbf{2020 Outburst}} \\
    \hline
    \multirow{1}{*}{3130010128} & 58966.566 & $14.77^{+0.09}_{-0.14}$ & $0.500^{+6.571}_{-0.500}$\tnote{*} &$0.57^{+1.10}_{-0.09}$ &  79.79 (111) \\
    \hline   
    \multicolumn{6}{c}{\textbf{2021 Outburst}} \\
    \hline
    \multirow{1}{*}{4130010101} & 59470.783 & $1.38^{+0.13}_{-0.11}$ & $1.630^{+0.075}_{-0.075}$ & $1.27^{+0.05}_{-0.05}$  & 89.04 (97)\\
    \hline
    \multirow{1}{*}{4130010102} & 59471.115 & $1.58^{+0.11}_{-0.09}$ & $1.585^{+0.054}_{-0.054}$ & $1.48^{+0.04}_{-0.04}$  & 118.81 (104)\\
    \hline
    \multirow{1}{*}{4130010103} & 59472.085 & $2.07^{+0.14}_{-0.12}$ & $1.639^{+0.046}_{-0.047}$ & $1.88^{+0.05}_{-0.05}$ & 162.76 (103)\\
    \hline
    \multirow{3}{*}{4130010104} & 59473.061 & $2.38^{+0.28}_{-0.18}$ & $1.750^{+0.102}_{-0.098}$ & $2.17^{+0.12}_{-0.10}$ &  103.09 (91)\\
    & 59473.382 & $2.45^{+0.16}_{-0.18}$ & $1.645^{+0.059}_{-0.059}$ & $2.19^{+0.07}_{-0.07}$  & 98.01 (98)\\
    & 59473.830 & $2.41^{+0.27}_{-0.19}$ & $1.634^{+0.088}_{-0.089}$ & $2.28^{+0.11}_{-0.11}$  & 118.10 (93)\\
    \hline
    \multirow{2}{*}{4130010105} & 59474.032 & $3.04^{+0.18}_{-0.20}$ & $1.820^{+0.057}_{-0.057}$ & $2.57^{+0.09}_{-0.08}$ & 137.68 (99)\\
    & 59474.485 & $3.16^{+0.15}_{-0.19}$ & $1.675^{+0.050}_{-0.052}$ & $2.89^{+0.08}_{-0.08}$ &  131.22 (100)\\
    \hline
    \multirow{1}{*}{4130010106} & 59475.246 & $4.28^{+0.21}_{-0.25}$ & $1.898^{+0.047}_{-0.048}$ & $3.70^{+0.10}_{-0.11}$ & 90.97 (104) \\
    \hline
    \multirow{3}{*}{4130010107} & 59476.084 & $4.45^{+0.27}_{-0.29}$ & $1.970^{+0.049}_{-0.050}$ & $4.02^{+0.12}_{-1.23}$ & 104.21 (102) \\
    & 59476.407 & $4.97^{+0.28}_{-0.35}$ & $2.030^{+0.020}_{-0.051}$ & $4.53^{+0.15}_{-0.14}$ & 89.73 (102)  \\
    & 59476.659 & $5.14^{+0.35}_{-0.36}$ & $2.047^{+0.054}_{-0.056}$ & $4.91^{+0.11}_{-0.16}$ & 94.27 (101) \\
    \hline
    \multirow{2}{*}{4130010108} & 59477.184 
    & $5.24^{+0.35}_{-0.36}$ & $2.090^{+0.049}_{-0.051}$ & $4.74^{+0.14}_{-0.15}$ & 87.76 (102) \\
    & 59477.562 & $5.91^{+0.29}_{-0.39}$ & $2.088^{+0.054}_{-0.053}$ & $5.43^{+0.19}_{-0.18}$ & 103.74 (102) \\
    \hline
    \multirow{3}{*}{4130010111} & 59480.793 & $14.29^{+0.16}_{-0.16}$ & $0.947^{+0.705}_{-0.947}$ & $5.32^{+3.82}_{-4.01}$ & 74.11 (104) \\ 
    & 59480.857 &$8.97^{+0.16}_{-0.14}$ & $2.291^{+0.033}_{-0.033}$ & $8.93^{+0.12}_{-0.12}$ & 80.76 (100) \\
    & 59480.923 & $6.95^{+0.11}_{-0.11}$ & $2.269^{+0.024}_{-0.035}$ & $6.90^{+0.09}_{-0.09}$ & 87.45 (97)  \\
    \hline
    \multirow{3}{*}{4130010112} & 59481.055 & $7.67^{+0.23}_{-0.21}$ & $2.221^{+0.055}_{-0.051}$ & $7.62^{+0.25}_{-0.15}$ & 85.14 (95) \\
    & 59481.515 & $5.81^{+0.23}_{-0.28}$ & $2.123^{+0.045}_{-0.047}$ & $5.18^{+0.15}_{-0.15}$ & 101.38 (104)  \\ 
    & 59481.898 & $6.23^{+0.31}_{-0.37}$ & $2.135^{+0.062}_{-0.063}$ & $5.44^{+0.21}_{-0.21}$ & 107.11 (97)\\
    \hline
    \multirow{1}{*}{4130010113} & 59482.803 & $7.16^{+0.42}_{-0.54}$ & $2.226^{+0.053}_{-0.052}$ & $6.54^{+0.23}_{-0.22}$ & 83.27 (101)\\ 
    \hline
    \multirow{3}{*}{4130010114} & 59483.255 & $7.55^{+0.46}_{-0.32}$ & $2.275^{+0.031}_{-0.032}$ & $7.79^{+0.07}_{-0.09}$ & 107.11 (97)\\
    & 59483.385 & $7.88^{+0.45}_{-0.66}$ & $2.238^{+0.060}_{-0.061}$ & $7.16^{+0.28}_{-0.28}$ &  93.35 (97) \\
    & 59483.648 & $14.19^{+0.19}_{-0.19}$ & $1.523^{+0.375}_{-1.281}$ & $6.69^{+3.03}_{-4.58}$ & 80.60 (103) \\
    \hline
    \multirow{1}{*}{4130010115} & 59484.416 & 
    $16.90^{+0.23}_{-0.23}$ & $1.644^{+0.360}_{-0.725}$ & $9.53^{+3.11}_{-4.24}$ & 82.61 (105)\\ 
    \hline
    \multirow{1}{*}{4130010116} & 59485.316 & $23.28^{+0.27}_{-0.27}$ & $1.491^{+0.678}_{-0.961}$ & $8.68^{+3.61}_{-6.25}$ & 90.22 (104)\\ 
    \hline
    \multirow{1}{*}{4130010118} & 59486.992 & $23.82^{+0.33}_{-0.27}$ & $1.753^{+0.884}_{-1.578}$ & $6.99^{+3.49}_{-6.19}$ & 63.52 (102)\\ 
    \hline
    \multirow{1}{*}{4130010119} & 59488.028 & $23.23^{+0.21}_{-0.56}$ & $1.534^{+0.485}_{-0.887}$ & $7.67^{+3.49}_{-4.03}$ & 61.15 (115)\\ 
    \hline
    \end{tabular*}
    \label{table3}

    \begin{tablenotes}
    \item[a] Two Gaussians with negative normalization included in the model to account for the absorption dips in the spectra
    \item[*] $\Gamma$ not well constrained. 
    \end{tablenotes}
    \label{spec_table}
  \end{threeparttable}
  
\end{figure*}

\subsubsection{2018 and 2020 Outburst}
The spectra corresponding to the QPO observation acquired during the 2018 outburst unveiled the presence of two discernible absorption dips centered at approximately 6.71 keV and 6.98 keV, with equivalent widths of 19.83 eV and 43.34 eV, respectively. These features exhibit characteristics akin to disk-wind-originated absorption lines, frequently encountered in black hole binaries residing in disk-dominated states \citep{king2014}.  To account for these features, two Gaussian components were incorporated into the spectral model. The addition of these Gaussians resulted in a substantial improvement in the goodness of fit (F-test probability $\sim$ 1.11$\times10^{-16}$). Moreover, the derived inner disk temperature of approximately 1.42 keV, combined with the normalization of the \texttt{DISKBB} component $\sim 200$, and a thermal fraction as high as 0.85, implies that the source was in a softer state in 2018 outburst obsevation. The predominance of thermal emission in this state, coupled with reduced sensitivity of $\emph{NICER}$ in the hard energy regime, may explain the unusually elevated value of the power-law index ($\Gamma$). In the 2020 outburst, no absorption dip was seen in the spectra of the QPO observations, and the photon index is very poorly constrained. 
This indicates a state dominated by soft emission, where the non-thermal component contributes only about 3\% to the total flux and thus $\emph{NICER}$ is unable to well-constrain the non-thermal parameters. 
The best-fit spectral parameters of these two observations are presented in Table \ref{spec_table}.

\subsubsection{2021 Outburst} 
\begin{figure*}
   \centering
    \begin{subfigure}[t]{0.48\textwidth}
        \includegraphics[width= 1\linewidth]{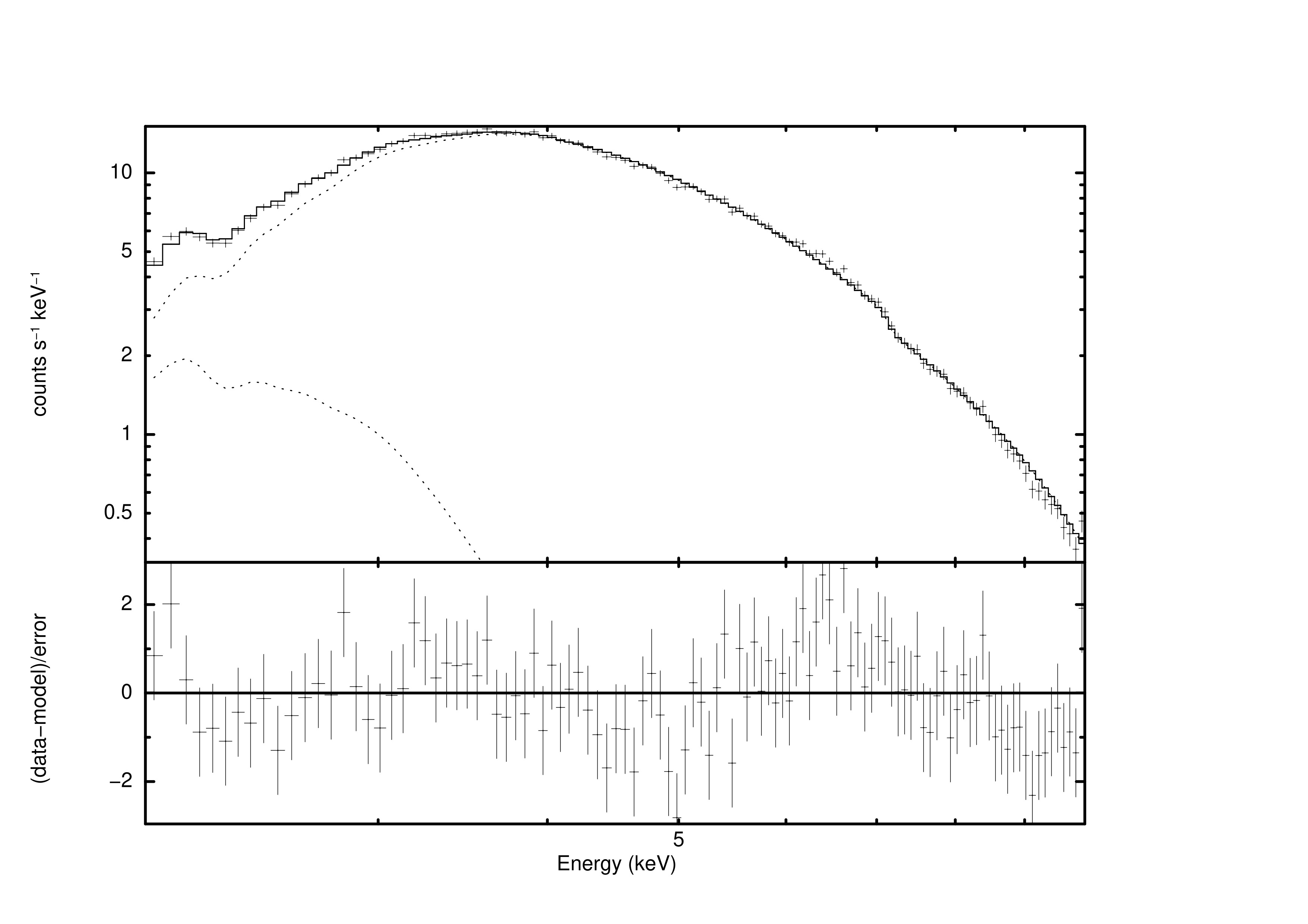}
        \caption{}
        \label{spec-a}
    \end{subfigure}
    \begin{subfigure}[t]{0.48\textwidth}
        \includegraphics[width=1\linewidth]{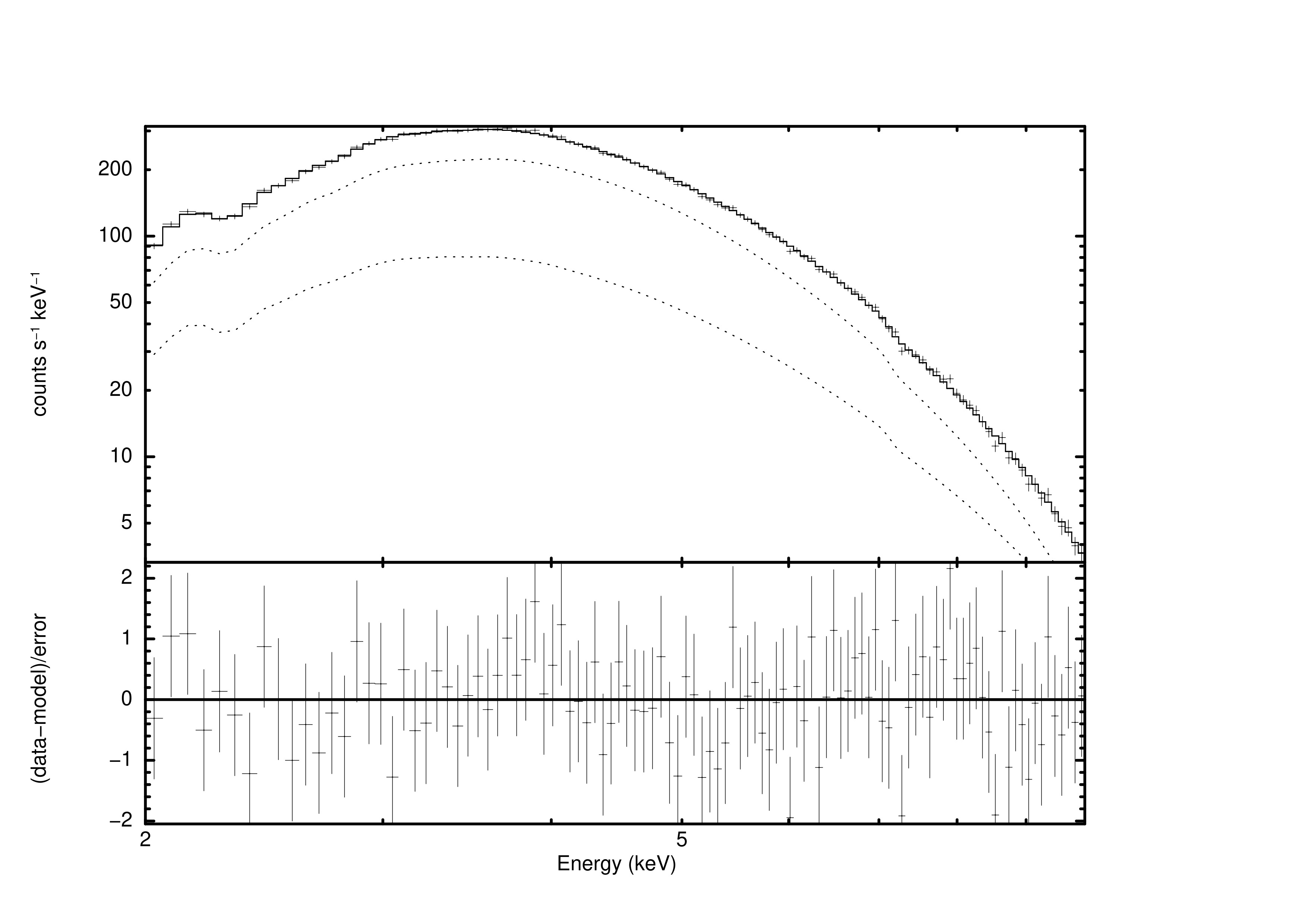}
        \caption{}
        \label{spec-b}
    \end{subfigure}

    \caption{Spectral fitting with \texttt{TBABS(POWERLAW+DISKBB)} model for OBSID 4130010102 (left) and OBSID 4130010118 (right) where \src was in the hard and soft state respectively during its 2021 outburst.}
    \label{fig:spec}
\end{figure*}

\begin{figure}[h]
    \includegraphics[width={1\linewidth}]{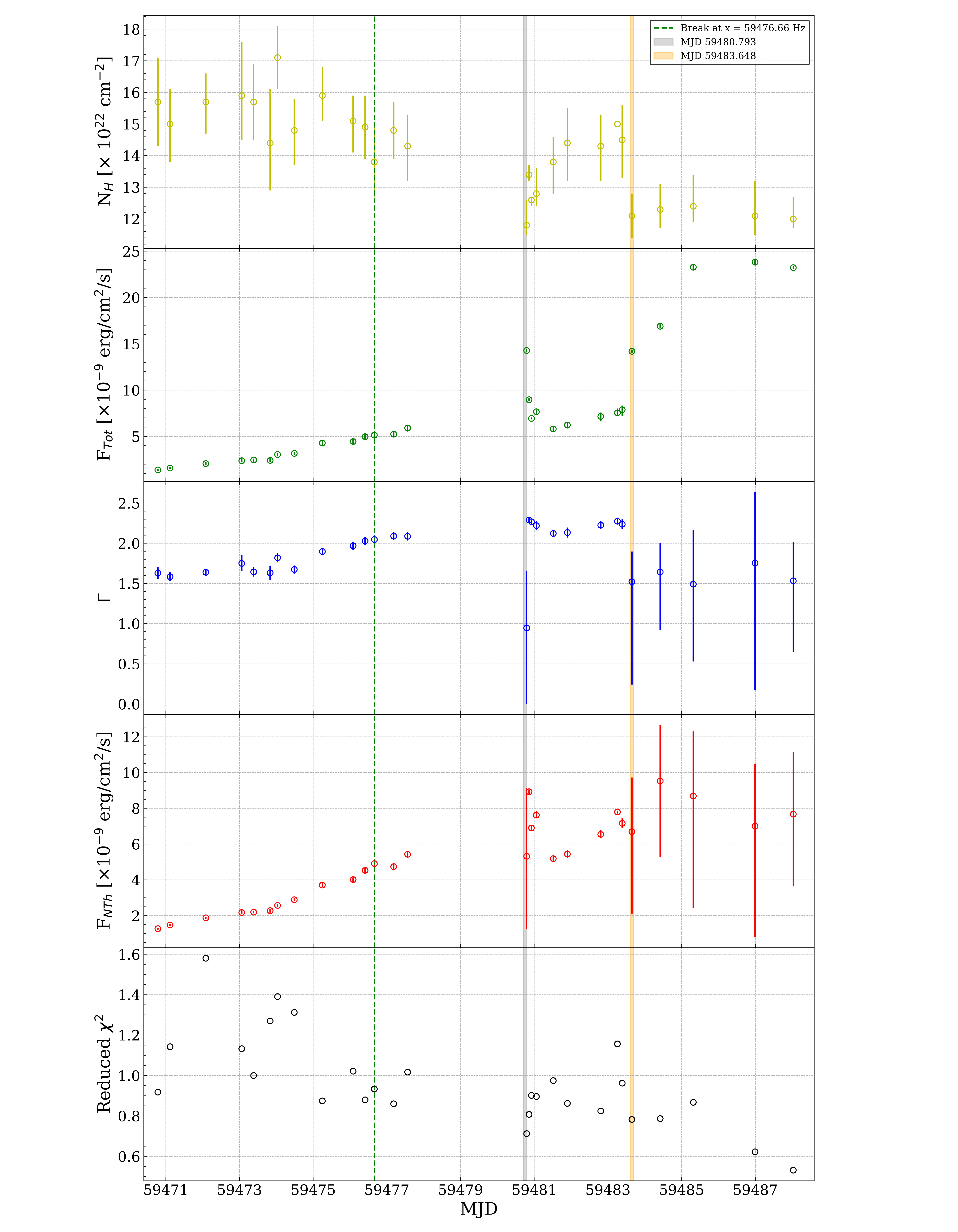}
    \caption{The time-evolution of primary spectral parameters, from top to bottom: the neutral Hydrogen density, the total flux (in 2.0--10.0 keV range), Power-law index $\Gamma$, non-thermal flux (in 2.0--10.0 keV range), and the reduced $\chi^{2}$ for the fit from the adopted spectral model \texttt{TBABS(POWERLAW+DISKBB)}. A noteworthy observation at MJD 59476.659 (green dotted line) marks a pivotal moment after which the relation between several parameters concerning QPO frequency changed (see Figure~\ref{break}). Regions marked by gray and orange shades indicate dates when the flux experiences sudden surges, leading to the disappearance of the QPO.} 
    \label{spec_params}
\end{figure}

For the 2021 outburst, we first analyzed individual single-day observations that featured either a single QPO or multiple QPOs. For solitary QPO cases, we used the {\tt NICERL3-spect} tool to generate the corresponding spectrum. However, for the cases with multiple QPOs, as shown in Figure \ref{OBSID4130010107}, where the QPO shows a significant evolution within OBSID 4130010107, we adopted a modified approach. To understand the unique timing and spectral properties inherent to the QPO at a particular time, just like the PDS, we also extracted the spectra separately for each segment over which the QPO frequency remained roughly constant
The spectra for these segments were generated using {\tt XSELECT}. However, background spectra and responses generated by {\tt NICERL3} were used in this analysis. 
We observed that the thermal parameters were unstable in the harder states, and for our analysis, we only considered the relatively stable spectral parameters. These derived spectral parameters, namely the power-law index ($\Gamma$), the total flux, and the nonthermal flux (in the 2--10 keV range), are presented in Table \ref{table3}. 
For soft states, where thermal parameters are well constrained, N$_{H}$ ($\sim$ 12 $\times$ 10$^{22}$ cm$^{-2}$) aligns well with the observations taken by \emph{Chandra} \citep{Pahari}. However, in observations in which the source was in the hard state, the estimated value of N$_{H}$ was relatively higher (12--17 $\times$ 10$^{22}$ cm$^{-2}$), which is close to the high values previously reported for \src \citep{Cui, wang2016}. However, in our analysis, it may also be attributed to the degeneracy of the thermal component and the effects of the absorption model are less discernible in the energy range selected for the spectral fitting of the \emph{NICER} data. 
Examples of spectral fitting with the model \texttt{TBABS(POWERLAW+DISKBB)} for hard and soft states are demonstrated in Figures~\ref{spec-a} and \ref{spec-b}, respectively. The temporal evolution of N$_{H}$, F$_{Tot}$, $\Gamma$, F$_{NTh}$ and the reduced $\chi^{2}$ is presented in Figure \ref{spec_params}.

\subsection{Critical frequency \texorpdfstring{$\nu_{c}$}{nu-c} and evolution of type-C QPOs} \label{result: critical frequency}
\begin{figure}
    \centering
    \begin{subfigure}{0.7\linewidth}
        \centering
        \includegraphics[width=1\linewidth]{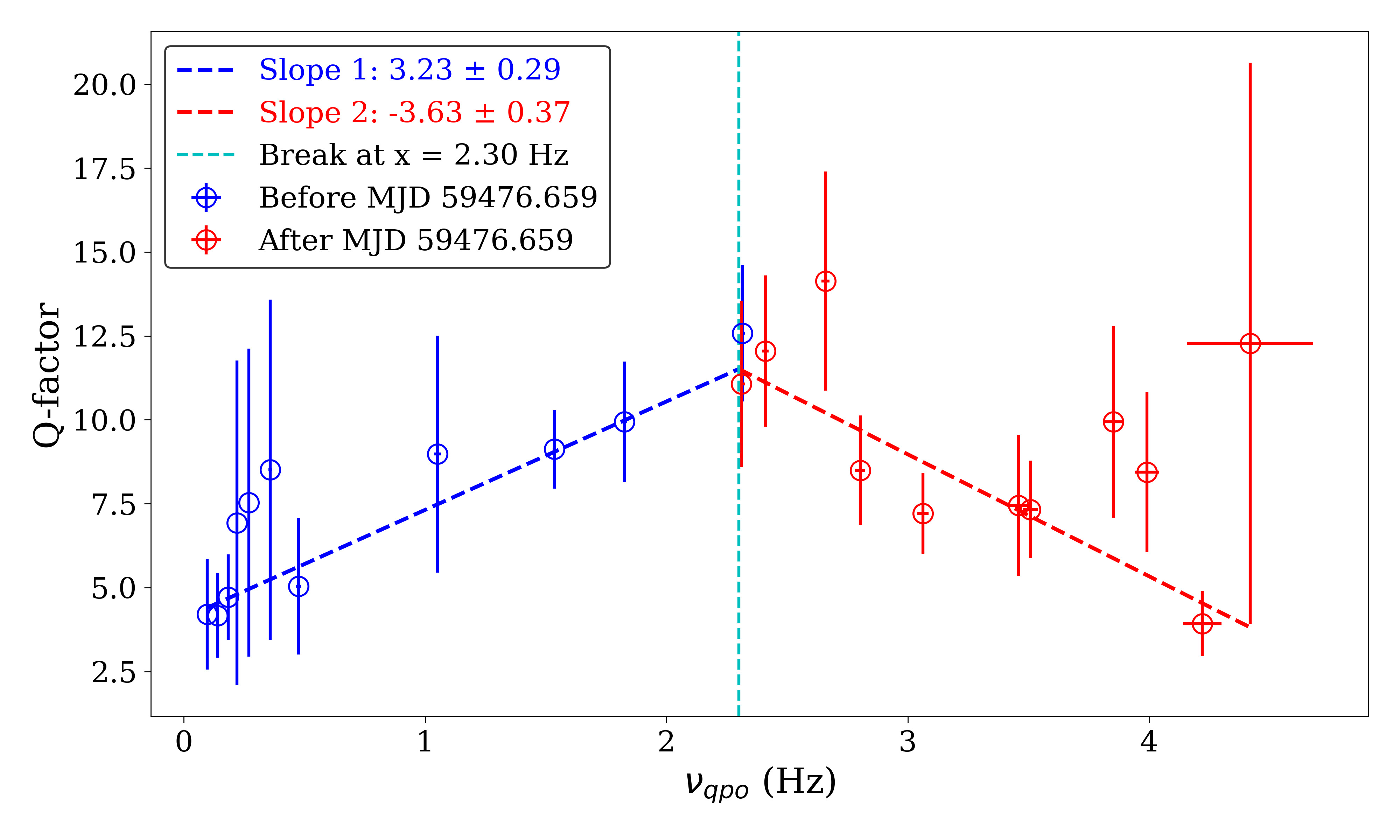}
        \caption{}
    \end{subfigure}
    \begin{subfigure}{0.7\linewidth}
        \centering
        \includegraphics[width=1\linewidth]{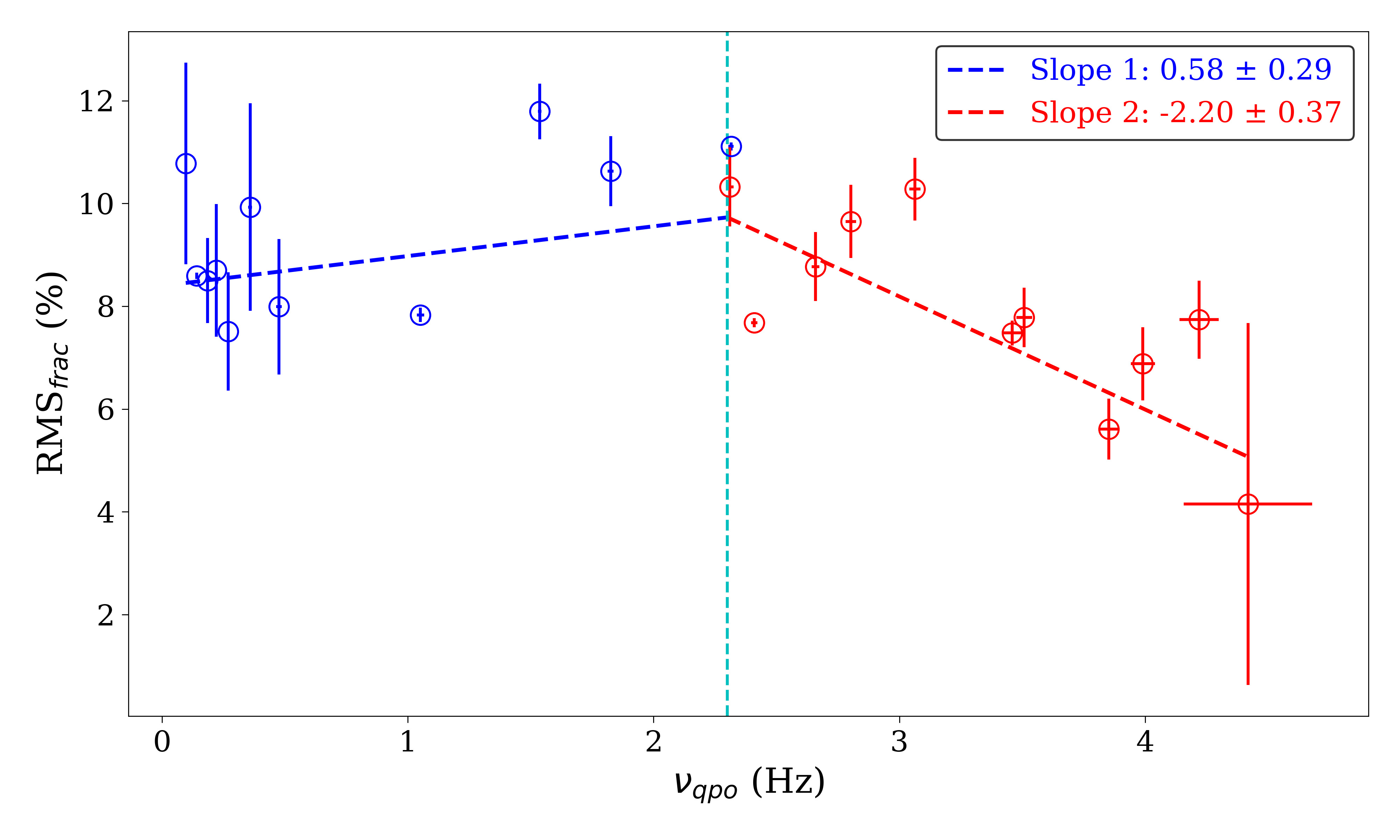}
        \caption{}
    \end{subfigure}
    \begin{subfigure}{0.7\linewidth}
        \centering
        \includegraphics[width=1\linewidth]{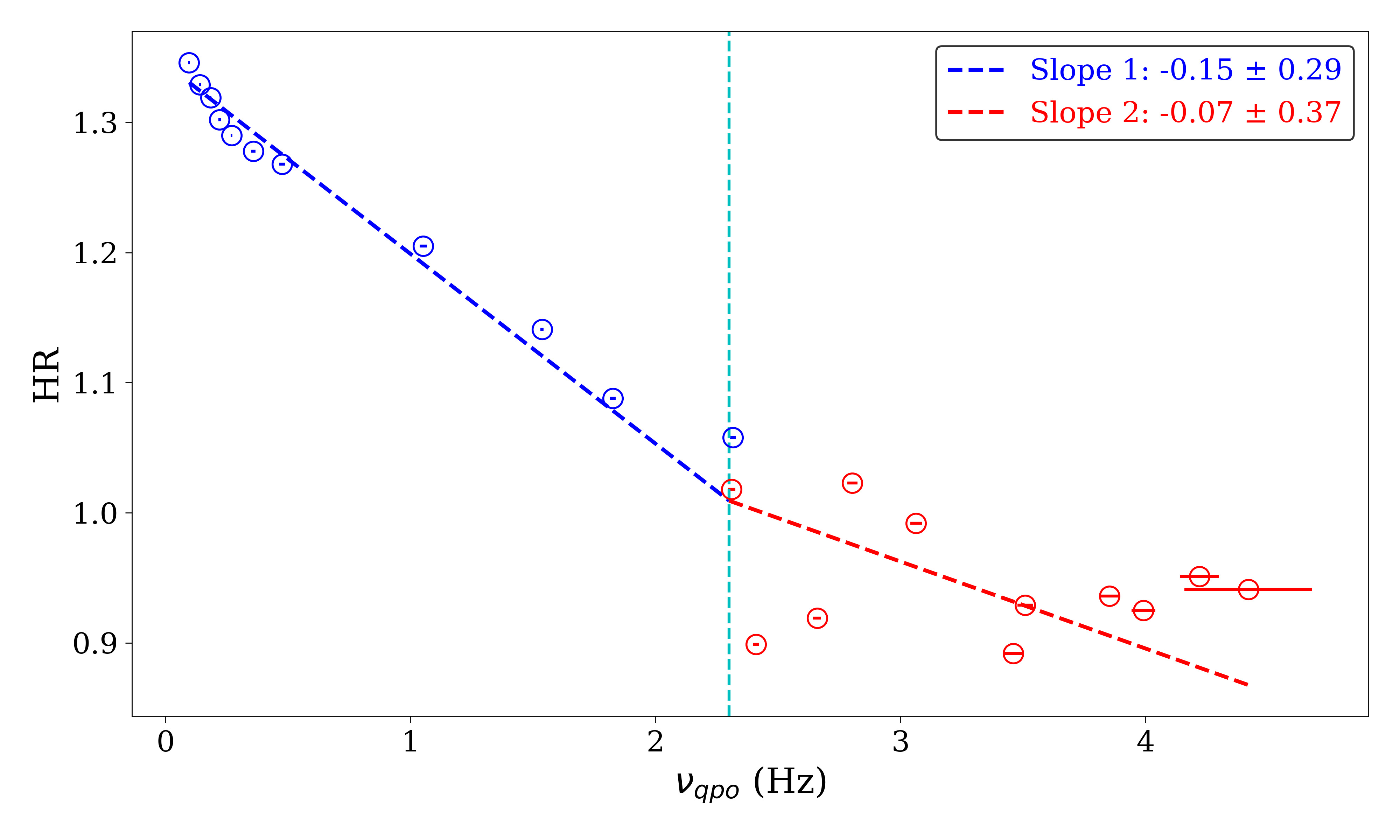}
        \caption{}
    \end{subfigure}
    \begin{subfigure}{0.7\linewidth}
        \centering
        \includegraphics[width=1\linewidth]{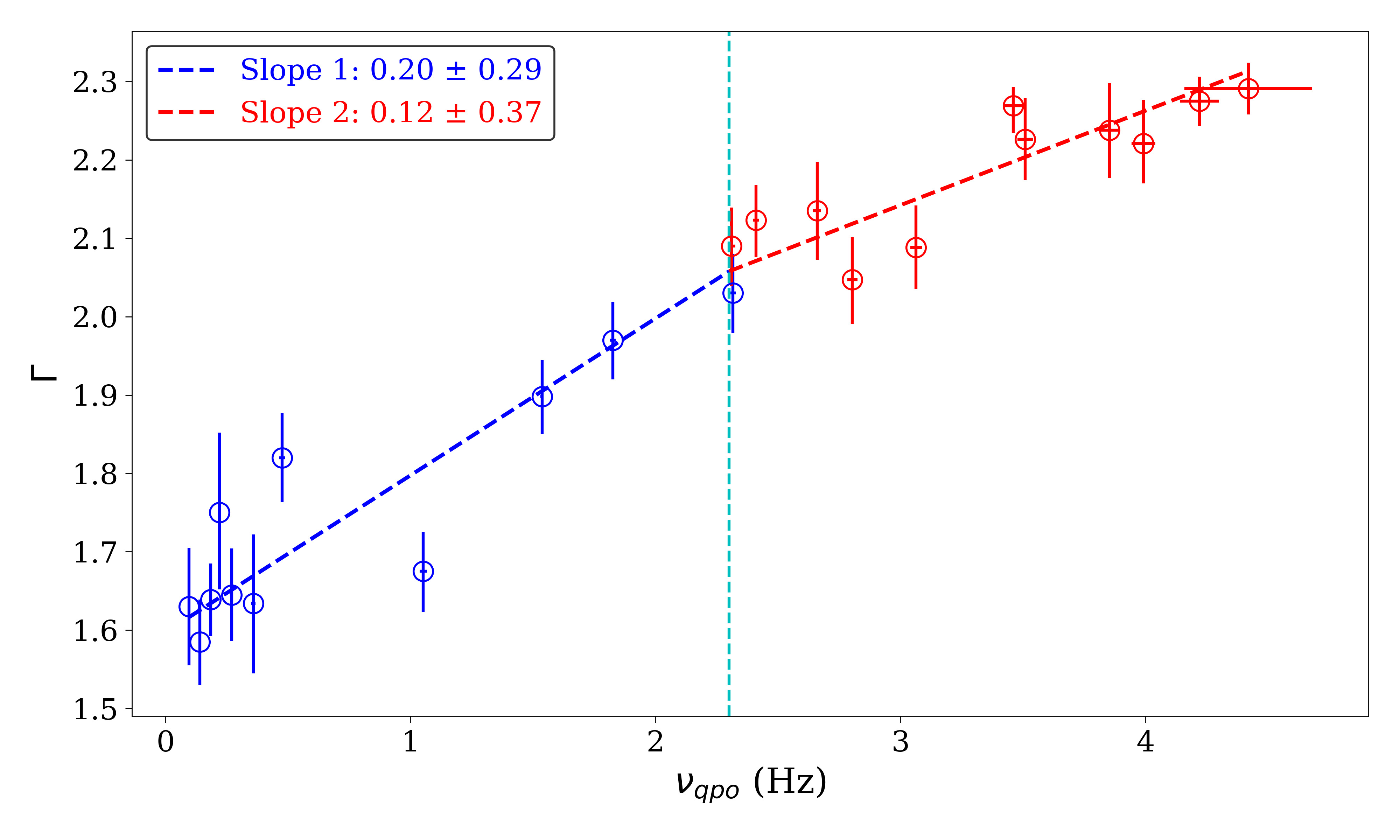}
        \caption{}
    \end{subfigure}
    \begin{subfigure}{0.7\linewidth}
        \centering
        \includegraphics[width=1\linewidth]{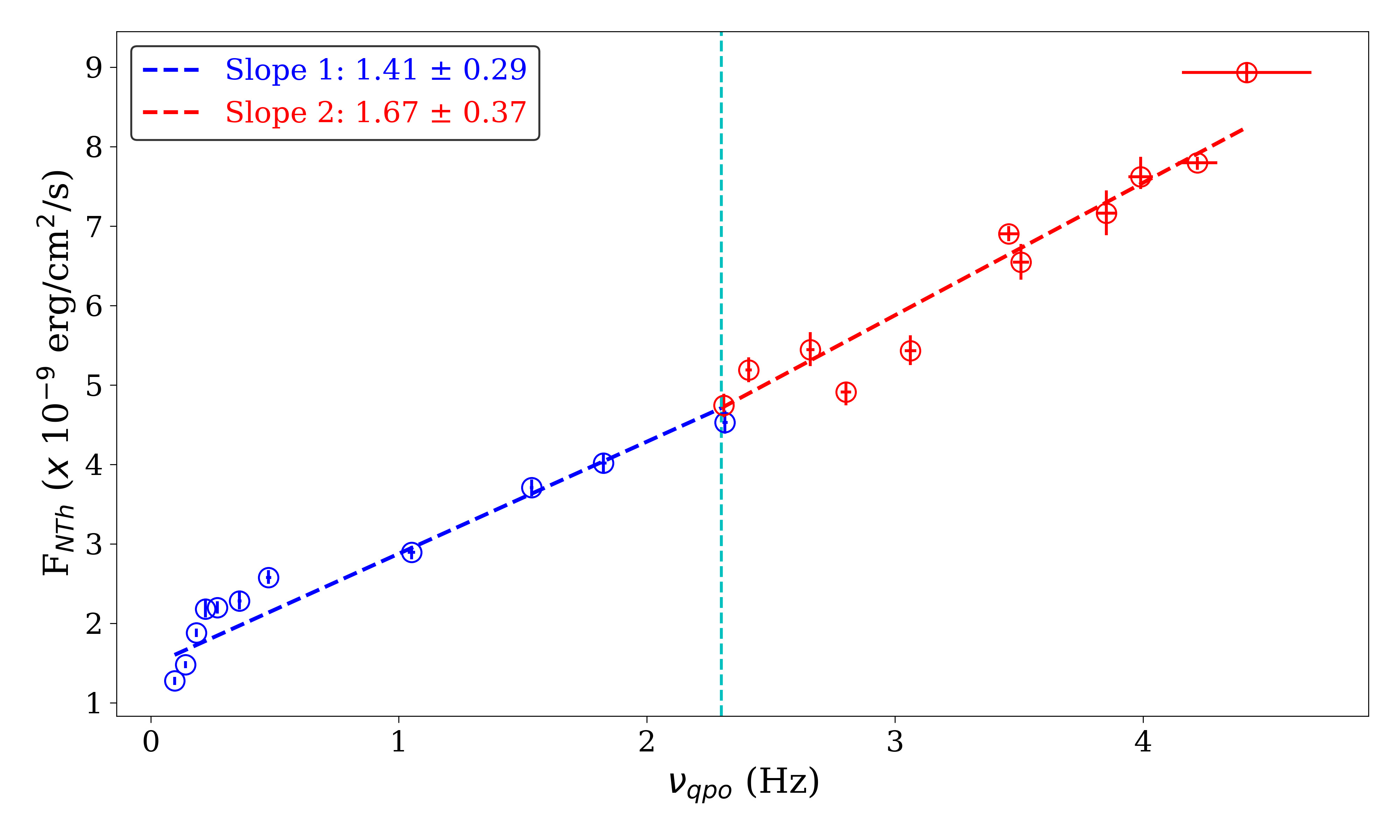}
        \caption{}
    \end{subfigure}    
    \caption{Dependence of Q-factor, $RMS_{frac}$, $HR$ (3.8--6.8 keV/2.0--3.8 keV), $\Gamma$, and the non-thermal flux (2.0--10.0 keV range) on frequency of the type-C QPOs $\nu_{qpo}$, observed for \src during its 2021 outburst. The blue and red points are the estimated values of the source parameters before and after MJD 59476.659. The best-fitted two-slope broken line is shown along with the estimated slopes (with 1$\sigma$ error) before and after the break. The vertical green dashed line shown at nearly 2.31 Hz represents the critical frequency, $\nu_{c}$.}
    \label{break} 
\end{figure}

For the QPOs observed in the 2021 outburst, we conducted an in-depth analysis of the intricate relationship between key spectral and timing model parameters and the type-C QPO frequency ($\nu_{qpo}$) observed when the source was in the hard spectral state.  
Interestingly, Figure \ref{break}(a) shows significant changes in the evolution of the Q-factor with respect to the QPO frequency, showing distinct behaviors below and above $\sim$2.31 Hz. There is a significant reversal in the slope of the two straight lines before and after the critical frequency ($\nu_{c}$). Notably, the evolution of the fractional RMS of the QPOs with $\nu_{qpo}$, as shown in Figure \ref{break}(b), displayed an indication of this break, declining beyond $\nu_{qpo} >$ 2.31 Hz. Similarly, the HR, depicted in Figure \ref{break}(c), experienced a variation as the QPO frequency increased, with a less steep correlation post $\nu_{c}$. However, it is crucial to note that the considerable error bars in HR values were not accounted for in the plot. 
As can be seen from Figure \ref{break}(d), a strong correlation was obtained between $\Gamma$ and $\nu_{qpo}$ that corresponded to a correlation coefficient of 0.97. We also observe that the slope for $\Gamma$-$\nu_{qpo}$ slightly evolves beyond $\nu_{qpo}$>2.31 Hz showing a noticeable but weak change. 
Correspondingly, as shown in Figure \ref{break}(e), following a QPO frequency of 2.31 Hz, the correlation between the QPO frequency and $F_{NTh}$ becomes a little stronger. By examining the combination of these trends in Figure \ref{break}, a pivotal transition point around $\nu_{c}$ ($\sim$ 2.31 Hz in this case) emerged, fundamentally reshaping the evolutions of both spectral and timing parameters.

\begin{figure}
  \centering
  \includegraphics[width=1.0\linewidth]{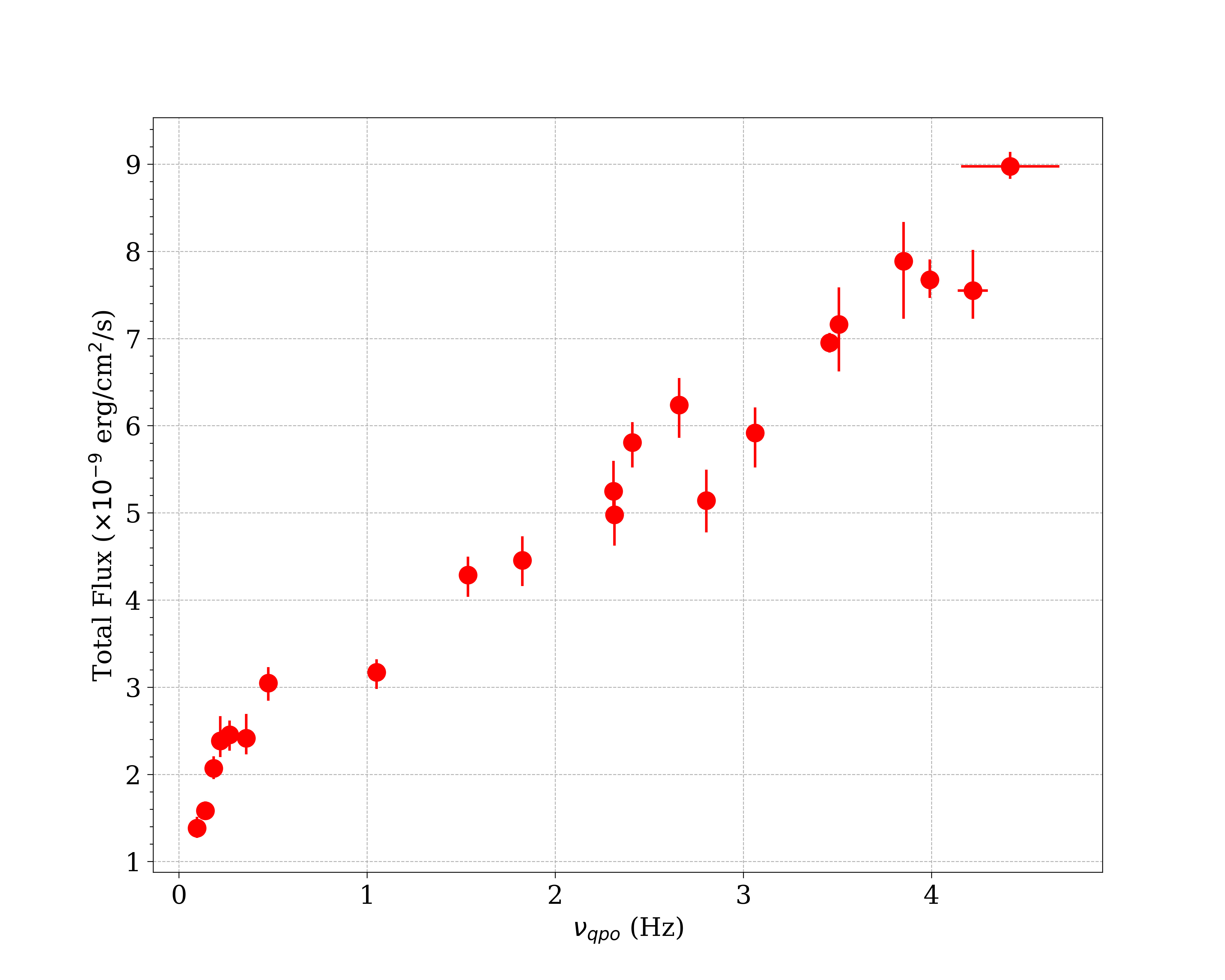}
  \centering
  \caption{\small Total Flux (in 2.0--10.0 keV range) as a function of type-C QPO frequency seen for \src during its 2021 outburst.} 
  \label{totflux}
\end{figure}

To deduce the critical frequency more accurately, we postulated a uniformity in the occurrence of the breakpoint across all the parameters in Figure \ref{break} as functions of $\nu_{qpo}$. 
We compared the evolution with a straight line (SL) and a two-slope broken line (BL) model and conducted an F-test to examine the requirement of the latter model with respect to the former.
Table~\ref{bpl} presents the $\chi^{2}$ values for the SL and BL fits, along with the chance probability of needing the BL model, estimated from the F-test.  
Notably, while the $F_{NTh}$ vs. $\nu_{qpo}$ relation did not exhibit a significant improvement with the BL model, the dependence of $\Gamma$, HR and RMS$_{frac}$ on $\nu_{qpo}$ is slightly improved when considering a BL model.  From the chance probability obtained via F-tests, it can be seen that the break was significantly detectable (beyond $3 \sigma$) only in the Q-factor$-\nu_{qpo}$ relation. Finally, we do not see any break in the case of evolution of the total flux with $\nu_{qpo}$, however, it is clear from Figure~\ref{totflux} that the type-C QPOs exhibit a very strong correlation with the total flux, corresponding to a Pearson's correlation coefficient of 0.99.

\begin{table}
    \centering
    \caption{Goodness of fit and F-test probabilities while fitting a straight line (SL) and a two-slope broken line (BL) line to the spectral and timing parameters with respect to $\nu_{qpo}$ for the type-C QPOs shown in Figure \ref{break}.}
    \small 
        \begin{tabular}{|c|c|c|c|}
        \hline
        Relation & SL   & Two-slope BL  &  F-test \\
         &  $\chi^{2}$ (dof)&$\chi^{2}$ (dof) & probability \\
        \hline
        Q-factor$-\nu_{qpo}$ & 50.20 (20) & 10.49 (19) & 6.96 $\times$ 10$^{-8}$ \\
        RMS$_{frac}-\nu_{qpo}$ & 1090.04 (20) & 832.78 (19) &  2.55 $\times$ 10$^{-2}$ \\
        HR$-\nu_{qpo}$ & 190.80 (20) & 159.85 (19) &  7.02 $\times$ 10$^{-2}$ \\
        $\Gamma-\nu_{qpo}$ & 24.62 (20) & 20.21 (19) &  5.59 $\times$ 10$^{-2}$ \\
        F$_{NTh}-\nu_{qpo}$ & 192.61 (20) & 182.58 (19) & 0.31 \\
        \hline
    \end{tabular}
    \label{bpl}
\end{table}

The detected critical frequency might also be associated with the final orbit observation on MJD 59476.659 represented by the green vertical line in Figure \ref{spec_params} (refer to the final orbit observation with OBSID 4130010107 in Table \ref{spec_table}). This is also shown in Figure \ref{OBSID4130010107} (red), which depicts the light curve and the PDS of this observation. 
Notably, $\Gamma$ remained consistently greater than 2 for all the QPO observations immediately after this epoch. 
This break could also indicate a transition from the LHS to the HIMS.

\subsection{Flare and transition to Soft Spectral state} \label{result: flare and transition}

During the ascending phase of the 2021 outburst, specifically on MJD 59480.793 and 59483.648 (refer to the gray and orange vertical bars in Figures \ref{qpo_parameters} and \ref{spec_params} and the gray and orange points in Figures \ref{lc2021} and \ref{hid2021}), $\emph{NICER}$ captured an intriguing phase characterized by a significant enhancement in the total flux (F$_{Tot}$). The power-law index ($\Gamma$) showed a sudden decrease which is consistent with the hardening of the spectra during these two observations, which can also be seen in the HID shown in Figure~\ref{hid2021}. It should also be noted that during both epochs, the type-C QPO vanished. 

\begin{figure*}[t]
   
    \begin{subfigure}[t]{0.33\textwidth}
        \includegraphics[width= 1\linewidth]{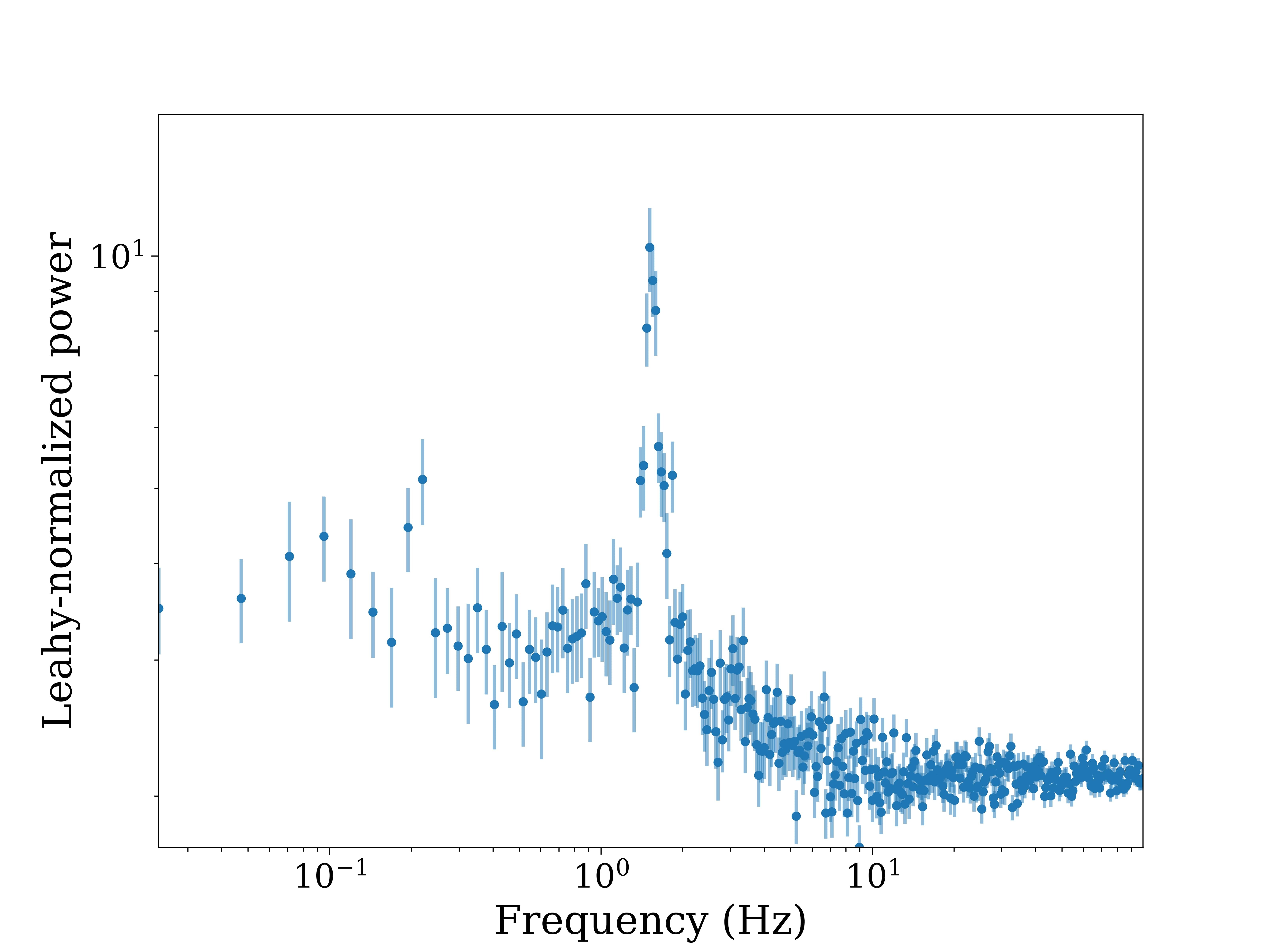}
        \caption{}
        \label{fig:pds-a}
    \end{subfigure}
    \begin{subfigure}[t]{0.33\textwidth}
        \includegraphics[width=1\linewidth]{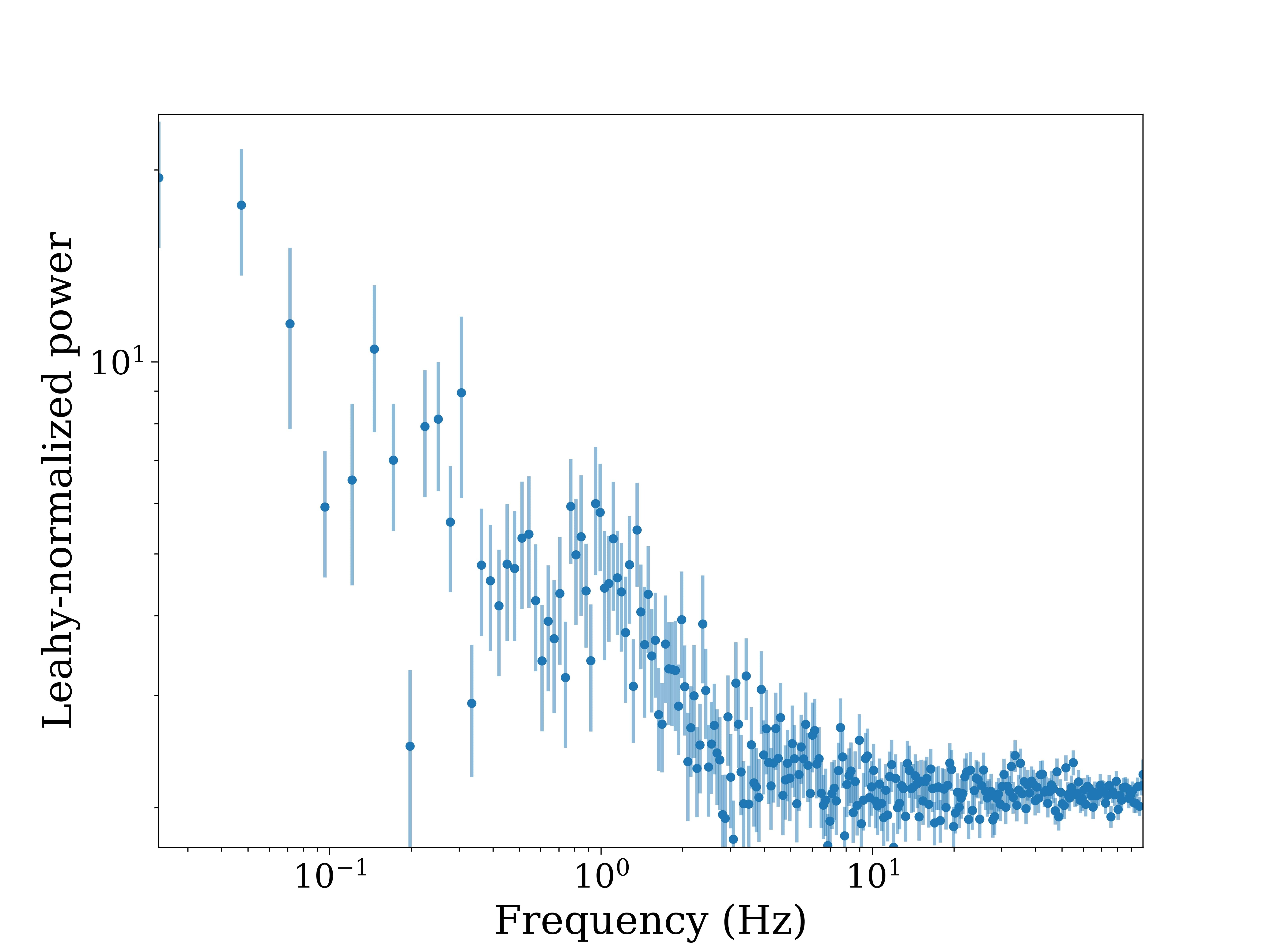}
        \caption{}
        \label{fig:pds-b}
    \end{subfigure}
    \begin{subfigure}[t]{0.33\textwidth}
        \includegraphics[width= 1\linewidth]{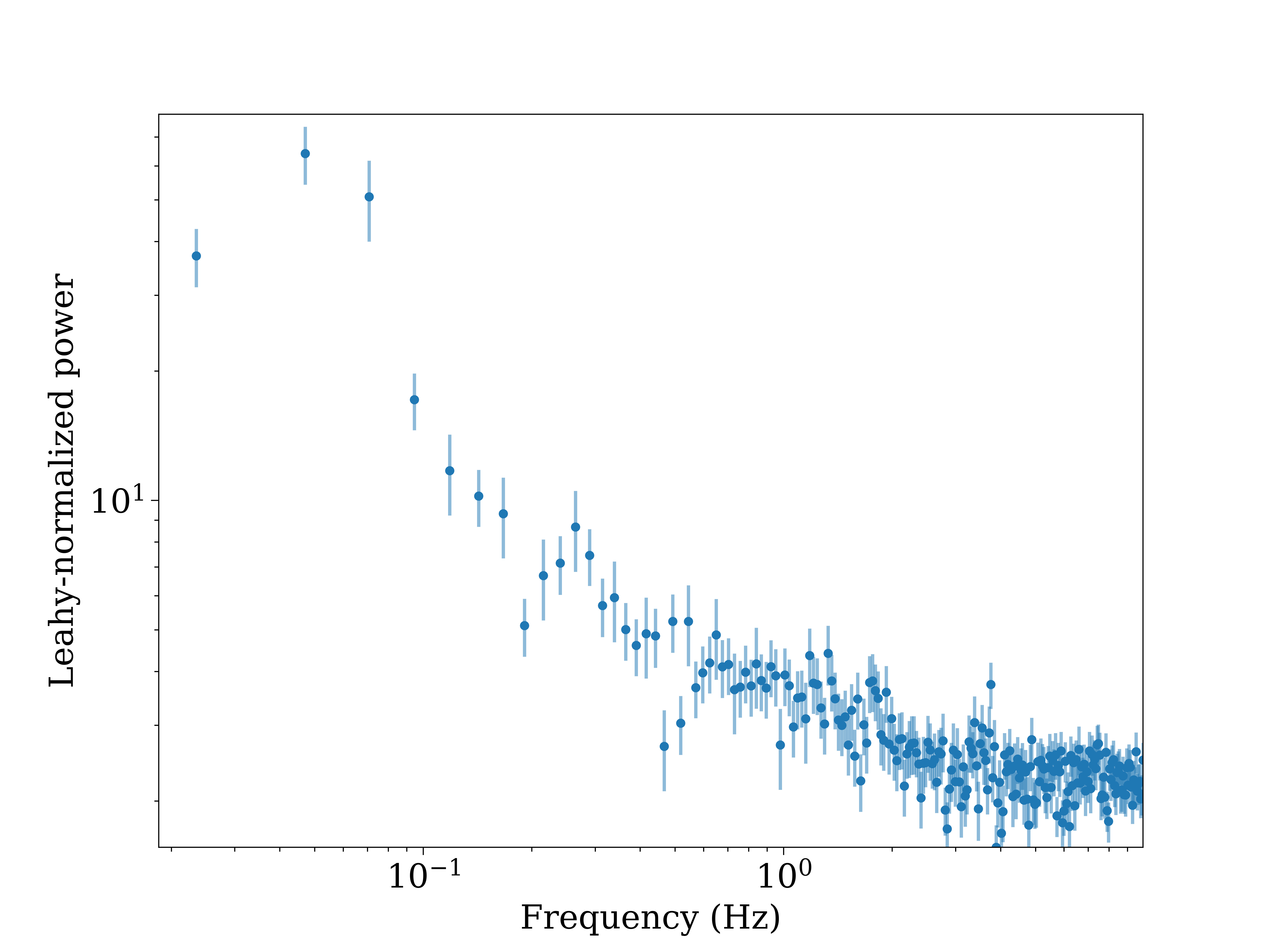}
        \caption{}
        \label{fig:pds-c}
    \end{subfigure}
    
    \caption{The PDS of the corresponding observations (a) type-C QPO, (b) the flaring event where the QPO is not visible, probably due to high count rates, and (c) the QPOs observed in the higher accretion regimes where we observe a peaked noise centered around $\sim$ 0.06 Hz (QRM) along with a QPO at $\sim$ 0.27 Hz.}
    \label{fig:both}
\end{figure*}

The observation taken on MJD 59480.793 displayed a distinctive signature compared to the overall evolution, which becomes pronounced when observing the total flux decrease from 14.29$_{-0.16}^{+0.16}\times 10^{-9}$ erg/cm$^{2}$/s to 8.97$_{-0.14}^{+0.16}\times 10^{-9}$ erg/cm$^{2}$/s in less than 90 minutes in the subsequent observation (OBSID 4130010111 in Table \ref{spec_table}). This implies that the event observed on MJD 59480.793 (see gray vertical line in Figures \ref{qpo_parameters} and \ref{spec_params}) was an X-ray flare. An equivalent change in the count rate from 614.74$\pm$2.03 counts/s to 360.61$\pm$1.45 counts/s can also be seen for the same observation in Figure~\ref{LC123-126}. Figure~\ref{fig:pds-b} depicts the PDS corresponding to this flaring event. 
The type-C QPOs reappeared post the flare on MJD 59480.857, similar to one shown in Figure \ref{fig:pds-a}. Subsequently, the total flux exhibited a fluctuation, initially decreasing and later increasing over a period of nearly three days (see Figures \ref{LC123-126} and \ref{spec_params}). The frequency of the type-C QPOs was seen to mimic these fluctuations. However, at the onset of the second event on MJD 59483.648, the total flux increased once more from 7.88$_{-0.66}^{+0.45}\times 10^{-9}$ erg/cm$^{2}$/s to 14.19$_{-0.19}^{+0.19}\times 10^{-9}$ erg/cm$^{2}$/s, leading to the subsequent disappearance of type-C QPOs. The effect of these two phenomena can also be seen in the HID, as is evident from the gray and orange points in Figure~\ref{hid2021}.

Interestingly, following this phase, the total flux displayed a continuous ascent from 14.19$_{-0.19}^{+0.19}\times 10^{-9}$ erg/cm$^{2}$/s to 23.23$_{-0.56}^{+0.21}\times 10^{-9}$ erg/cm$^{2}$/s, propelling the source into softer states. Associated with this transition, $\Gamma$ and the non-thermal flux showed large uncertainties from MJD 59483.648 onward (see Figure~\ref{spec_params}). During this period, type-C QPOs were replaced by a different kind of QPOs with frequencies ranging from 0.27 to 0.77 Hz (as indicated in Table \ref{table1}). The corresponding PDS of one of these observations taken on MJD 59484.416 (OBSID-4130010115) can be seen in Figure~\ref{fig:pds-c}. This indicates that the event observed at the start of MJD 59483.648 was not an X-ray flare, but rather a transition from HIMS to a higher accretion state. These QPOs, observed in these softer states, featured weaker intensities with fractional RMS values below 3\%. Since the RMS values of these QPOs were low, we performed an F-test to verify that these features were not random noise. The F-test revealed that the introduction of an additional Lorentzian significantly improved the fit, with significance levels of 3.67$\sigma$, 2.67$\sigma$, 3.22$\sigma$, and 2.64$\sigma$ for the QPOs observed in OBSIDs-4130010115, 4130010116, 4130010118, and 4130010119, respectively. These results indicate that these four QPOs were less significantly detectable, but at least two of them satisfied a $3~\sigma$ detection threshold even using a complementary approach. Moreover, it should be noted that the QPOs detected in the final three OBSIDs reported in Table \ref{table1} occur consistently around the similar frequency of $\sim$0.7 Hz across different days, providing further support that the detection is likely not random. 
This weaker feature coexisted with a pronounced noise peak at even lower frequencies (0.06--0.20 Hz), as exhibited by the PDS in Figure \ref{fig:pds-c}. The last four rows of Tables \ref{table1} and \ref{table3} provide a detailed breakdown of the timing and spectral parameters of these observations. It should be noted
that these weak QPOs were not included in the subsequent analysis on the correlation between various temporal and spectral parameters of type-C QPOs shown in Figure~\ref{bpl} and discussed in Section $\S$\ref{discussion: timing and spectral params}.
A detailed discussion on these weak QPOs is given in Section $\S$\ref{discussion: flares and transitions}.

%% file: discussion.tex
\section{DISCUSSION} \label{discussion}
In this study, we conducted a detailed spectro-timing analysis of \src during its three distinct outbursts in 2018, 2020, and 2021, utilizing the extensive observational data obtained from \emph{NICER}. It is clear from Figures \ref{lc2018} and \ref{lc2020} that the 2018 and 2020 outbursts predominantly exhibited characteristics typical of the High/Soft State (HSS), with X-ray flux remaining relatively stable and showing minimal variability. The PDS derived from the observations of the 2018 and 2020 outbursts by \emph{NICER} during the HSS only revealed one transient QPO each at frequencies below 0.3 Hz (check Table~\ref{table1}). Since these short-lived  QPOs were only seen in one observation of each outburst, we interpret that these unstable modulations are atypical of any other known forms of variability that can arise in these systems. Secondary outbursts, as observed in the 2018 and 2020 light curves, are common in BHXBs. \citet{Tomsick2000} detected LFQPOs during the decay phase of the 1998 outburst of \src. However, no QPOs were detected during the Soft to Hard state transition in the 2018 and 2020 outbursts observed by \emph{NICER}. The absorption features seen at $\sim$ 6.71 keV and 6.98 keV during the QPO observation taken for the outburst in the year 2018 most likely emanated from Fe xxv and Fe xxvi ions within a low-velocity ionized disk wind \citep{king2014}. These moderately blue-shifted absorption lines have also been identified in the recent outburst of \src \citep{rawat2023b, Ankur}. 

After analysing the individual orbit data of the 2021 outburst from \emph{NICER}, however, we identified 21 type-C QPOs (see, for example, Figure \ref{fig:pds-a}) between MJD 59470.783 and 59483.255 when the source was in the Low/Hard and Hard-Intermediate state. Subsequently, our investigation unveiled substantial variations in the QPO frequency with the intensity throughout the course of a single-day observation on MJD 59476 (see Figure~\ref{OBSID4130010107}), motivating the extraction of PDS from light curves of individual orbits of the observation. We also noticed that the type-C QPOs abruptly disappeared during the third observation on MJD 59483.648 (refer Table \ref{table1} or see orange vertical line in Figure \ref{qpo_parameters}) and were later replaced by a different set of QPOs (see Figure~\ref{fig:pds-c}) whose properties are not entirely consistent with the properties of either type-A or type-B QPOs. In the following sections, we discuss the dependence of QPO parameters and various associated correlations.

\subsection{Dependence of QPO parameters on spectral parameters during the 2021 outburst} \label{discussion: timing and spectral params}

The type-C QPOs observed until the end of MJD 59483 displayed a significant positive correlation with the non-thermal photon index $\Gamma$ (refer to Figure \ref{break}(d)), exhibiting a strong Pearson correlation coefficient of 0.97 ($p$-value$\sim$10$^{-13}$). \citet{vignarca2003} reported a similar tight positive correlation between the QPO frequency and photon index for multiple BHXRB sources, with the correlated behaviour showing a turn-off at high QPO frequencies for certain sources. Such a correlation can be supported by the QPO at a particular frequency being related to a specific radius within the disk and subsequently undergoing Comptonization by the corona.
Moreover, from Figure \ref{erps}, the increasing amplitude of these QPOs with rising photon energy provides additional confirmation of their connection to the comptonizing region or the corona. 
An even stronger correlation of $\nu_{qpo}$ was seen with the total flux with the Pearson correlation coefficient going as high as 0.99 with corresponding $p$-value of $\sim$10$^{-17}$ respectively. Assuming the mass accretion rate scales with the total flux, this result further indicates a connection between the $\nu_{qpo}$ and the corresponding mass accretion rate. The LFQPOs have consistently exhibited a robust correlation with the mass accretion rate across various sources. A notable example is the 2017-18 outburst of GRS 1915+105 \citep{Liu2021}, where LFQPOs showcased a frequency evolution from 2.62 to 4.38 Hz, concurrent with an analogous rise in the mass accretion rate.

The observed relation between RMS$_{frac}$ and $\nu_{qpo}$ during the rising phase, as depicted in Figure \ref{qpoVSRMSVScr}, bears intriguing similarities with the findings for the source H 1743-322 reported by \citet{zhang2023} in a comparable frequency range of 0.33--9.43 Hz. \citet{zhang2023} examined various instances of type C QPOs spanning different outbursts from H1743--322 and observed that in the rising phase, the fractional RMS displays a mild elevation at frequencies below 1 Hz, followed by a relatively stable pattern around 1–2 Hz, and a pronounced decline at higher frequencies. Moreover, these patterns of QPO fractional RMS in relation to frequency remained consistent across different outbursts \citep{zhang2023}. In our analysis of the 2021 outburst of \src, we observed no QPOs with frequencies exceeding 4.4 Hz. Consequently, we could not adequately observe the decline in the QPO RMS-frequency pattern. Additionally, the absence of observations during the decay phase of the outburst further prevents the extension of this analysis to that phase.
\citet{zhang2023} proposed the Lense-Thirring (L-T) precession model \citep{ingram2019,stella} as one of the possible explanations for the typically observed evolution of $\nu_{qpo}$ and the intrinsic QPO RMS. In this framework, a variable coronal geometric shape (i.e., $h/r$) triggered by the frame-dragging effect influences the wobbling frequency and intrinsic amplitude of the QPO. The dynamic aspect of the L-T precession model \citep{ingram2009} relates $\nu_{qpo}$ to the outer radius, $r$, of the corona, indicating that increasing $\nu_{qpo}$ implies a decreasing $r$. However, \citet{Liu2021} did not observe this anti-correlation, suggesting that the L-T precession model might not align with their case, despite having seen a positive correlation between $\nu_{qpo}$ and the mass accretion rate. In the literature, the estimated inner disk radius values corresponding to the observed LFQPOs were often observed to be a few times $R_g$ (gravitational radius), which is highly inconsistent with the truncation radius expected from the Lense-Thirring precession model involving an inner hot flow \citep{yang2022, Nathan2022}. Furthermore, \citet{Zhao2024} reported the first polarimetric study of LFQPOs in BH LMXBs where they observed the absence of modulation in the polarization degree and the polarization angle with the QPO phase in contrast to strong modulation of the photon index with QPO phase. This poses further challenges to the processing hot inner flow model.

Recent models have provided significant insights into the mechanisms driving type-C QPOs in black hole X-ray binaries. 
\citet{Ma2021} explained their observations of LFQPOs above 200 keV in the black hole LMXB MAXI J1820+070  using a jet precession model, which attributes the QPOs to the relativistic precession of a small-scale jet as opposed to a precessing hot inner flow.
\citet{Karpouzas2020} proposed a Comptonization framework considering a spherical corona to reproduce the evolutionary characteristics of the QPOs. In this model, they do not explicitly assume any origin of the QPO frequency but suggest oscillatory modes in the corona as a possible origin.
Using this model, the frequency dependence of the phase lag of the type-C LFQPOs in the black hole LMXB GRS 1915+105 and the transition of the phase lag at a critical frequency of $\sim$1.8 Hz could be successfully explained \citep{Karpouzas}. A similar variable Comptonization model has also been implemented by \citet{Bellavita} to generate the energy-dependent RMS and phase lag behavior of both type-B and type-C LFQPOs.
In the context of our results, the variable Comptonization model aligns well with the observed energy dependence of RMS in type-C QPOs. Since we could not detect significant phase lags in our analysis, a detailed comparison of the phase lag behaviour in the light of this model could not be conducted. However, we observed that the photon index related to the Comptonization had a strong correlation with the QPO frequency. 
Furthermore, in our case, the break in the evolution of certain spectral and timing parameters seen at a frequency of $\sim$2.31 Hz could also signal a switch in the dominant physical mechanism driving the observed QPO properties \citep{Karpouzas}. 
 
When considering the influence of the mass accretion rate as the primary determinant of the evolution of type-C QPOs, we particularly note the absence of QPOs with frequencies exceeding 4.4 Hz in our observation sample. 
Specifically, during the occurrence of the second flare-like event on MJD 59483.648 (refer to OBSID 4130010114 in Table \ref{table3}), a substantial surge in the total flux was observed, ascending from 7.88$\times10^{-9}$erg/cm$^{2}/s$ to 14.19$\times10^{-9}$erg/cm$^{2}$/s. Subsequently, the total flux entered a range unsupportive for generating type-C QPOs with frequencies surpassing the 4.4 Hz threshold. 
Figures \ref{fig:pds-a} and \ref{fig:pds-b} clearly show how the PDS change when type-C QPO is present, and when the mass accretion rate crosses the threshold value necessary to sustain these periodicities. This observation further shows the remarkable sensitivity of type-C QPOs to variations in the mass accretion rate, underscoring the intricate relationship between these two parameters. The type-C QPOs, therefore, are either not stable at higher accretion rates, or the disk does not remain thin \citep{Yorgancioglu} and as a result, the fluctuations in the non-thermal flux become hard to observe.

\subsection{Flares and transition between QPOs} \label{discussion: flares and transitions}
An additional captivating revelation emerged from the \emph{NICER} observation of the 2021 outburst, wherein the QPO exhibited a peculiar behavior. It is evident from Table~\ref{table1} that during the first orbit observation taken on MJD 59480.793, the QPO evolution abruptly halted, and it seemingly vanished from detection (also see the gray vertical line in Figures \ref{qpo_parameters} and \ref{spec_params}). 
This intriguing event coincided with a significant rise in the total flux and a decrease in $\Gamma$ (see OBSID 4130010111 in Table \ref{table3}). Following this sudden disappearance, the QPO reappeared as the total flux came down in subsequent observations taken on MJD 59480 and continued to evolve for three additional days, culminating with the third observation on MJD 59483 (refer to orange vertical line in Figures \ref{qpo_parameters} and \ref{spec_params}), during which the total flux again surged along with a lowering of $\Gamma$ making the source transition to higher accretion states. This transition is also evident from the light curve and HID shown respectively in Figure~\ref{lc2021} and \ref{hid2021} (black to orange transition). Subsequently, the total flux continued to rise until MJD 59489, leading to the emergence of distinct, albeit weaker, QPOs with an even lower frequency. 

In their comprehensive analysis, \citet{yang2022} meticulously studied the 2021 outburst of \src utilizing \emph{Insight}-HXMT data, during which they successfully identified type-C QPOs within the frequency range of $\sim$1.6--4.2 Hz. Notably, they also uncovered mHz Quasi-regular Modulations (QRMs) around 60 mHz, persisting from MJD 59483.785 to 59485.240. These QRMs displayed Q-factor values between 2 and 4 and fractional RMS amplitude of $\sim$11-16\% (measured within the ME 8--35 keV energy range). In contrast, their observations, which were quasi-simultaneous with \emph{NICER} during the intervals between MJD 59483.785 and 59485.909, did not yield any reported QPO occurrences. Further, to explore potential peaks within the 1--10 Hz frequency span, \citet{yang2022} included an additional Lorentzian component in the PDS, but did not significantly detect any low-frequency QPOs. However, in our analysis, the contemporaneous observations conducted by \emph{NICER} during the same period further enriched the dataset. The \emph{NICER} observation taken on MJD 59484.416 showcased the same QRM ($\sim 60$ mHz) as detected by \citet{yang2022}, along with a weak QPO centered around 0.27 Hz (Figure \ref{fig:pds-c}). The fractional RMS of the mHz QRM reported by \citet{yang2022} displayed an increase at higher energies, implying a strong correlation with the corona. These QRMs demonstrated substantial sensitivity to variations in the accretion rate, implying a robust interplay between these parameters. In the literature, mHz QRMs and accompanying QPOs were reported for \src at flux levels of approximately $14.0\times10^{-9}$ erg/cm$^{2}$/s in the 3--20 keV energy range, further emphasizing the intricate link between the phenomenon and accretion rate \citep{yang2022, trudolyubov2001, dieters}. In the \emph{NICER} energy range of 2.0--10.0 keV, the total flux on MJD 59484.416 was $\sim$ 16.9$\times$10$^{-9}$ erg/cm$^{2}$/s. \emph{NICER} observations of the source over the next three days displayed no QRM, which further provides evidence of the fact that QRMs are observed only in a specific flux range. Further analysis of these three observations from MJD 59485 to 59488 yielded a QPO with a frequency ranging from 0.67 to 0.77 Hz (refer Table~\ref{table1}) along with broad noise at even lower frequencies (around 0.2 Hz). This broad noise may correspond to the QPO observed on MJD 59484.416 at $\sim$0.27 Hz, which seems to manifest in subsequent observations as a feature around $\sim$0.2 Hz in these subsequent PDS. It seems that the QRM was replaced by a broad noise in the PDS in these subsequent observations, again highlighting their sensitivity to higher total flux and consequently higher mass accretion rate. Furthermore, our analysis unveiled a correlation between the frequency of these weak QPOs and the total flux (likely mimicing the mass accretion rate variation), as detailed in Tables~\ref{spec_table} (last four observations).

Remarkably, the fractional RMS associated with these QPOs remained below 3\%, exhibiting pronounced prominence within the \emph{NICER} energy band of 3.0--6.0 keV. Following the classification criteria outlined by \citet{casella2004}, the Q-factor $\ge$6 suggests a classification as type-B QPOs. Conversely, the fractional RMS value falling below or equal to 3\% would typically indicate type-A QPOs. However, it is important to note that the observed frequencies of these QPOs, which are less than 1 Hz, align more closely with the characteristics associated with type-C QPOs. But, unlike type-A QPOs, which typically do not last long, these QPOs persisted for about four days.
Though type-B QPOs were found at low frequencies in the range of $\approx$ 1--3 Hz by \citet{motta2011}, it still does not lie in the frequency range in which the four above-mentioned QPOs were detected. Such transitions from type-C to type-B have been previously seen to take place after the occurrence of an X-ray flare in a black hole LMXB MAXI J1820+070 by \citet{Homan}. In their study, the X-ray flare was followed by a radio flare, which is also proposed to be the cause of the type-B QPOs. No radio observations were made for the 2021 outburst of \src, thus hindering the confirmation of these QPOs as type-B QPOs. As a consequence, the connection between these QPOs and the radio jet could not be tested. Thus, the results obtained from our analysis can not necessarily classify these QPOs as one QPO or the other at this current juncture because the properties of these QPOs do not align with any known type of QPO. However, the persistence of these QPOs and their occurrence around the peak of the outburst are possible indicators of the type-B classification.
Additionally, it should be noted there remain some questions related to the significant presence of these QPOs. 
However, complementary F-test statistical analysis reveals that the chances of these noises being generated randomly are low, as the addition of an extra Lorentzian component improves the chi-square of the fit significantly in most cases, reducing the weighted residuals around the QPO frequency. This is further supported by the significance levels derived from the F-test (see $\S$\ref{result: flare and transition}) along with the significance of these QPOs as mentioned in Table~\ref{table1}. Furthermore, these QPOs meet our considered criterion of having Q-factors $\geq$ 2, a standard threshold for distinguishing noise from QPOs, and are thus retained in our analyses.

\subsection{Breaks in parameters as a function of \texorpdfstring{$\nu_{qpo}$}{nu-qpo}}
Our investigation revealed a breakpoint in the evolution of both spectral and timing parameters with the type-C QPO frequency, occurring at approximately 2.31 Hz (see Figure~\ref{break}). This distinct transition was evident in the relationship between the Q-factor and $\nu_{qpo}$.
The presence of this breakpoint suggested a change in the physical processes governing these quantities and prompted us to probe further into similar trends. Interestingly, \citet{rawat2023a} indicated similar trends at $\sim3$ Hz in MAXI J1535--571. Notably, these transitions were pronounced in QPO lags, corona size, and seed photon temperature. We also found parallels with the work of \citet{zhang2020} in GRS 1915+105, where a QPO frequency of around 2 Hz coincided with a shift in QPO lags from Soft to Hard. Remarkably, both these instances correlated with times when radio jets were suppressed. On the other hand, no radio observations were reported for the 2021 outburst of \src to further compare our findings with \citet{zhang2020} and \citet{rawat2023a}.
In the study by \citet{rawat2023a}, a dual corona model was proposed to explain the observed behaviors. In this model, a smaller corona in close proximity to the black hole dominated time-averaged spectra, while a more extensive corona, potentially a jet, dominated lag spectra. 
We observed a stronger correlation between $\Gamma$ and the QPO frequency before the critical frequency with the corresponding slope of the straight line as 0.20$\pm$0.29. However, this correlation became less steep for higher $\nu_{qpo}$ (see Figure~\ref{break}(d)), with the corresponding slope of the fitted straight line changing to 0.12$\pm$0.37. This trend is consistent with \citet{rawat2023a} at low QPO frequencies but extends to even lower frequencies ($<$1 Hz) in our study. It is worth noting that \citet{rawat2023a} used {\tt NTHCOMP} and {\tt DISKBB} models to fit the non-thermal and thermal components of the spectra, where they evoked the dual corona model to explain the discrepancy of the temperature at the inner disk radius with the seed photon temperature for the comptonization. The degeneracy observed in the thermal parameters in the hard state observations limits us to further investigate such relations in this work.

Further, we noted a decrease in the strength of QPOs beyond the critical frequency (Figure~\ref{break}b), again in alignment with observations by \citet{rawat2023a}. This reduction in QPO strength seems to be associated with the diminishing vertical extent of the corona. The observed, albeit weak ($\sim 2\sigma$), existence of break in the relationship between $\Gamma$ and $\nu_{qpo}$ around 2.3 Hz or after MJD 59476.659 resulted in a weakened correlation between these parameters. This change signified a reduced frequency dependence on the photon index of the non-thermal coronal emission. Moreover, the consistency observed in the evolution of the non-thermal flux, $F_{NTh}$ (see Figure~\ref{break}e) given that the indication of the break was much weaker in this case, suggested stabilization of the evolution of the coronal geometry and indicated that $\nu_{c}$ could simply be related to the change in the spectral state from Low/Hard to Hard-Intermediate state. 
However, it is important to acknowledge a potential limitation in our analysis. The proficiency of \emph{NICER} in capturing data within higher energy bands remains somewhat constrained. Consequently, an intrinsic caveat exists in interpreting the precise dynamics between $F_{NTh}$ and $\nu_{qpo}$, particularly within the framework of higher energy ranges. As such, while the observed trend is certainly intriguing, it is recommended to approach this specific correlation with a nuanced perspective and an awareness of this inherent limitation.

%% file: conclusion.tex
\section{CONCLUSION}
We have conducted an extensive investigation of the variability of \src using \emph{NICER} data during its outbursts in the years 2018, 2020, and 2021 and detected 27 low-frequency QPOs, with 25 observed in the 2021 outburst alone. Out of these, 21 were type-C QPOs detected in the Low/Hard and Hard-Intermediate states. We further carried out a spectro-temporal study of the detected QPOs, and obtained a strong correlation between the type-C QPO frequency, the total flux, non-thermal flux, and the non-thermal photon index ($\Gamma$) during the rising phase of its outburst in the year 2021. 
Moreover, we observed several correlations of the spectral parameters (non-thermal flux, $\Gamma$) and the timing parameters ($RMS_{frac}$, Q-factor) with the QPO frequency, out of which the Q-factor-$\nu_{qpo}$ relation showed a significant break at the critical frequency, $\nu_{c} \sim$ 2.31 Hz. This result provides valuable insights into the underlying physical processes governing these systems, opening avenues for further exploration into the complex interplay of accretion and the QPO phenomenon. 

It was particularly intriguing to observe the transition of the QPOs and their properties alongside the change from low to high accretion regimes during the same outburst. The QRM detected at $\sim$ 60 mHz in between this transition suggests a possible relation between the QRM and the properties of the weak QPOs detected by \emph{NICER} in the softer states. Comparing our findings with those in previous work on type-C QPOs, certain results were found to be consistent with the predictions of either the Lense-Thirring precession model or the variable Comptonization model.
However, the significant breaks observed at the critical frequency lack a clear explanation, which can be addressed through further theoretical and observational investigations.